\documentclass[3p]{elsarticle}
\usepackage[dvips]{epsfig}

\usepackage{amsmath}
\usepackage{amssymb}
\usepackage{bm}
\usepackage[mathscr]{eucal}
\usepackage{theorem}
\usepackage{graphicx}
\usepackage{psfrag}
\usepackage{algorithm}
\usepackage{algpseudocode}
\usepackage{color}
\usepackage{wasysym}
\include{graphix}
\usepackage{framed}
\usepackage{enumitem}%
\usepackage{lipsum}
\usepackage{float}
\usepackage{MnSymbol}
\usepackage{cancel}
\usepackage{soul}
\usepackage{multirow}
\usepackage{caption,subcaption}
\usepackage{mwe}
\usepackage{url}

\usepackage{tikz}
\usetikzlibrary{shapes}
\usepackage{xspace}

\newcommand{\typical}{\overline{\ell}}

\newcommand{\vecx}{\mathbf{x}}
\newcommand{\vecy}{\mathbf{y}}
\newcommand{\vecz}{\mathbf{z}}

\newcommand{\vecb}{\mathbf{b}}
\newcommand{\vecrhs}{\mathbf{f}}
\newcommand{\vecrhsT}{\mathbf{d}}

\newcommand{\vecf}{\mathbf{f}}
\newcommand{\vecg}{\mathbf{g}}

\newcommand{\mxA}{\mathbf{A}}

\newcommand{\mxL}{\mathbf{L}}
\newcommand{\mxR}{\mathbf{R}}
\newcommand{\mxU}{\mathbf{u}}
\newcommand{\mxV}{\mathbf{v}}

\newcommand{\mathd}{\mathrm{d}}

\newcommand{\myqed}{\nobreak \ifvmode \relax \else
      \ifdim\lastskip<1.5em \hskip-\lastskip
      \hskip1.5em plus0em minus0.5em \fi \nobreak
      \vrule height0.75em width0.5em depth0.25em\fi}

\begin{document}
\title{A Batched GPU Methodology for Numerical Solutions of Partial Differential Equations}
\author[mymainaddress]{Enda Carroll}
\author[mymainaddress]{Andrew Gloster}
\author[mymainaddress]{Miguel Bustamante}
\author[mymainaddress]{Lennon \'O N\'araigh\corref{mycorrespondingauthor}}
\cortext[mycorrespondingauthor]{Corresponding author}
\ead{lennon.onaraigh@maths.ucd.ie}

\address[mymainaddress]{School of Mathematics and Statistics, University  College Dublin, Belfield, Dublin 4, Ireland}

\date{\today}

\begin{abstract}
In this paper we present a methodology for data accesses when solving batches of Tridiagonal and Pentadiagonal matrices that all share the same left-hand-side (LHS) matrix.  The intended application is to the numerical solution of Partial Differential Equations via the finite-difference method, although the methodology is applicable more broadly. By only storing one copy of this matrix, a significant reduction in storage overheads is obtained, together with a corresponding decrease in compute time. Taken together, these two performance enhancements lead to an overall more efficient implementation over the current state of the art algorithms cuThomasBatch and cuPentBatch, allowing for a greater number of systems to be solved on a single GPU. We demonstrate the methodology in the case of the Diffusion Equation, Hyperdiffusion Equation, and the Cahn--Hilliard Equation, all in one spatial dimension.  In this last example, we demonstrate how the method can be used to perform $2^{20}$ independent simulations of phase separation in one dimension.  In this way, we  build up a robust statistical description of the coarsening phenomenon which is the defining behavior of phase separation. We anticipate that the method will be of further use in other similar contexts requiring statistical simulation of physical systems.

\vspace{0.1in}
\noindent{\bf{Program Summary}} \\
Program Title: CUDA Batched Tridiagonal and Pentadiagonal Schemes \\
Licensing Provision: Apache License 2.0 \\
Programming Languages: C, C++, CUDA \\
Computer: Variable, equipped with CUDA capable GPU \\
Operating System: Linux, Mac and Windows \\
Nature of Problem: Various implementations of batched Pentadiagonal and Tridiagonal solvers exist for CUDA using an interleaved data layout format.
All of the current state of the art implementations require large amounts of memory due to the need for every thread to have its own copy of the $\mxA$ matrix.
There are many situations, particularly in the batch solving of PDEs in 1D and 2D, where the $\mxA$ matrix is the same, thus the data overheads are unnecessarily large reducing the number of systems that could be solved on a single GPU.\\
Solution method: In this paper we eliminate the $\mxA$ matrix requirements and allow every thread to access the $\mxA$ matrix, a dramatic saving in memory and also grants an additional speed-up over the existing state of the art.  \\
Tridiagonal Functions Source: \url{https://github.com/EndCar808/cuThomasConstantBatch} \\
Pentadiagonal Functions Source: \url{https://github.com/munstermonster/cuPentConstantBatch} \\

\end{abstract}

\maketitle

\noindent Keywords: CUDA, GPUs, Tridiagonal, Pentadiaognal, Matrix Inverse, C, C++

\section{Introduction}
\label{intro}
In this paper we consider batched numerical solutions of systems of the form $\mxA \vecx = \vecb$ on a GPU, in particular where $\mxA$ is a tridiagonal or pentadiagonal matrix.
Batched solutions of such tri/pentadiagonal matrix systems are becoming increasingly prevalent methods for tackling a variety of problems on GPUs which offer a high level of parallelism~\cite{zhang2010fast, kim2011scalable, cuThomasBatch, gloster2019cupentbatch}.

The intended application of the methodology is to the numerical solution of partial differential equations via finite-difference methods.  We therefore showcase the application of these methods to the Diffusion Equation, the Hyperdiffusion Equation, and the Cahn--Hilliard Equation.  We focus on examples in one spatial dimension, although higher-dimensional problems can be tackled using this same methodology, together with an Alternating Direction Implicit (ADI) discretization scheme~\cite{ADIGPU, gloster2019custen}.  The Diffusion Equation is one of the most-studies equations in Computational Science, and we use this to showcase the method and to quantify its performance.  Similarly, we use the Hyperdiffusion Equation as a benchmark problem for a pentadiagonal solver.  However, in this article, we focus in much more detail on the last example, namely the one-dimensional Cahn--Hilliard equation.  In this example, we demonstrate the computational power of the new methodology by characterizing the phenomenon of coarsening and phase-separation in a statistically robust fashion.

We emphasize that the methodology in this article may find application wherever GPUs are useful for accelerating established computational procedures.
For example, in Fluid Mechanics, there has been a broad application of GPUs where solutions of Poisson's equation are commonly required~\cite{hockney1964fast, valeroPoisson}, tsunami modelling and simulation~\cite{reguly2018volna}, numerical linear algebra~\cite{laraVariableBatched, HaidarSmallMatrices}, batch solving of 1D partial differential equations~\cite{gloster2019cupentbatch}, ADI methods for 2D simulations~\cite{ADIGPU, gloster2019custen} and modelling mesoscopic-scale flows using Lattice--Boltzmann methods~\cite{pedro_lbm}.
Also, in the field of gravitational wave data analysis, much work is done simplifying complex waveforms and analyses to achieve results in a physically reasonable and desirable amount of time. 
A large portion of this work involves using cubic splines for interpolation~\cite{splineIEEE}, a highly parallelizable process that has shown promising results using GPUs.  Yet further examples of areas using GPUs include image in-painting \cite{imageInpainting}, simulations of the human brain~\cite{cuHines, BrainsOilGas}, and animation~\cite{pixar}.  Consequently, the methodology presented in this article may find very broad applicability, although for definiteness we focus on the numerical solutions of certain canonical partial differential equations for the present purposes.

In the remainder of this Introduction, we place the present methodology in the context of the state-of-the-art.  At the library level various implementations/functions have been developed for batched solves, examples include the Batched BLAS project \cite{dongarra2017optimized, dongarra2017design}, the CUDA libraries cuBLAS, cuSPARSE and cuSOLVER, and the Intel MKL library.
In this paper we will focus on the development of tridiagonal and pentadiagonal batch solvers with single $\mxA$, multiple $\vecb$, matrices in CUDA for application in solving batches of 1D problems and 2D ADI methods.
CUDA, an API extension to the C/C++ language, allows programmers to write code to be executed on NVIDIA GPUs.
In particular we develop solvers in this work which are more efficient in terms of data storage than the state of the art, this saving of data usage is due to the fact we have only one global copy of the $\mxA$ matrix rather than one for each thread/system.
While the primary improvement is in data storage reduction the new functions also provide increased speedup, due to better memory access patterns, when compared to the existing state of the art.
This increase in efficiency is also beneficial as it leads to further savings in resources, both in terms of run-time and electricity usage.
The latter of which is of increasing concern as modern supercomputers become larger, requiring more and more electricity which has both financial and environmental implications.
The need for both energy efficient hardware and algorithms is becoming an ever more prevalent trend.

To date all existing batch pentadiagonal solvers in CUDA require that each system in the batch being solved have its own copy of the $\mxA$ matrix entries, leading to a sub-optimal use of device memory (GPU memory) in terms of unnecessary space usage and increased memory access overhead when only one global copy of $\mxA$ is actually needed. We also note that, while there are options for single $\mxA$, multiple $\vecb$, tridiagonal and pentadiagonal matrices in cuSPARSE, these options overwrite the LHS data leading to a necessary correction step having to be performed when using these in iteration resulting in additional computational cost. Thus we propose the implementation of two solvers, one implementing the Thomas Algorithm for tridiagonal matrices and the other implementing the pentadiagonal equivalent as presented in~\cite{gloster2019cupentbatch, numalgC}, each with a single global copy of the $\mxA$ matrix and multiple $\vecb$ matrices. We will benchmark these implementations against cuThomasBatch \cite{cuThomasBatch} (implemented as gtsvInterleavedBatch in the cuSPARSE library) and existing work by the authors cuPentBatch \cite{gloster2019cupentbatch}, for tridiagonal and pentadiagonal matrices respectively, which are the existing state of the art for the algorithms we are interested in. We point the user to the papers \cite{cuThomasBatch} and \cite{gloster2019cupentbatch} for existing benchmark comparisons of these algorithms with multiple $\mxA$ matrices with the cuSPARSE library and comparisons with serial/OpenMP implementations. Thus we can compare our new implementations to the existing state of the art, cuThomasBatch and cuPentBatch, from which relative performance compared to the rest of cuSPARSE can be interpolated by the reader. 


The paper is laid out as follows. 
In Section~\ref{method} we outline the modified methodology of interleaved data layout for $\vecb$ vectors and a single global copy for the $\mxA$ matrix.
In Sections~\ref{tri} and~\ref{pent} we show how this methodology can be applied to  tridiagonal and pentadiagonal matrices respectively.  We include here benchmarks against the state-of-the-art algorithms to show performance is at least as good, if not better in most cases of batch size and matrix size. 
In the later parts of the paper, we move on to a specific physical application, namely the coarsening phenomenon in the 1D Cahn--Hilliard equation.  As such, in Section~\ref{1D} we present a numerical method for solving multiple copies of the 1D Cahn--Hilliard equation in a batch using an implicit numerical algorithm.  We show how this approach can be used to generate a statistically robust description of coarsening dynamics in a phase-separating liquid.  We continue with this study in Section~\ref{flowpattern}, where we show that with the same approach, a detailed parameter study of the forced Cahn--Hilliard equation (with a rich parameter space) can be explored using batch simulations to generate large amounts of data.  We again characterize the data with with statistical approaches, this time using k-means clustering to demarcate distinct regions of the parameter space.
Finally we present our conclusions in Section~\ref{con}.


\begin{figure}[htb]
	\centering
		\includegraphics[width=0.6\textwidth]{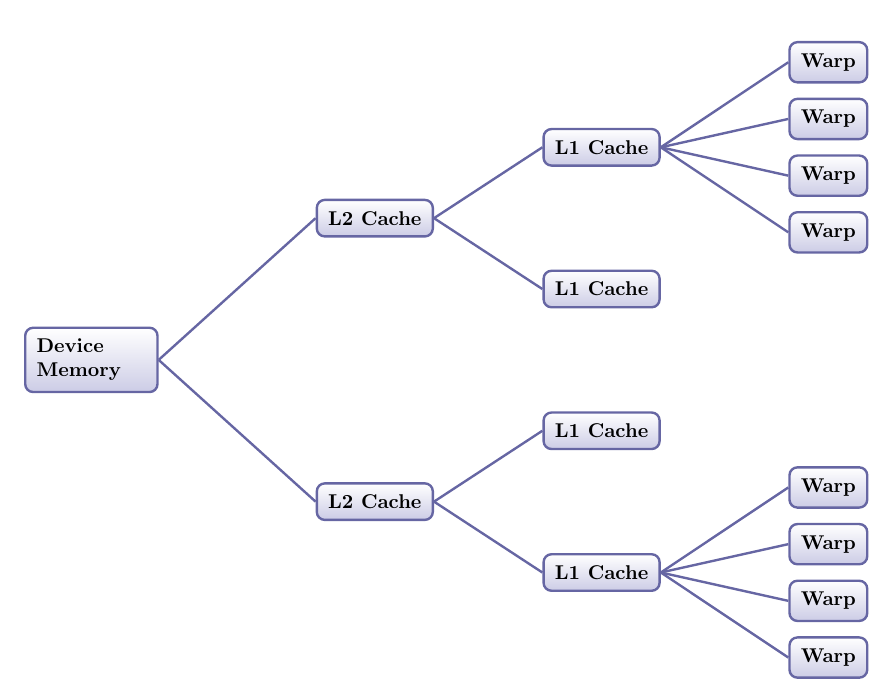}
\caption{Diagram of memory hierarchy in an NVIDIA GPU. We omit nodes for the purposes of the diagram, the top and bottom branches are representative of the memory routing from Device Ram to warp. When every thread is accessing the same memory location warps which share the same SM will receive an L1 cache hit, SMs sharing an L2 cache will receive an L2 cache hit and if the data is not present in the L2 cache it will be retrieved from the Device memory. Thus it can be seen here how the first warp to request a location in memory in a given group will be the slowest but accesses for following warps become faster and faster as the memory propagates down the tree into various caches.}
\label{memoryaccess}
\end{figure}

\begin{figure}[htb]
\centering
\includegraphics[width=0.4\textwidth]{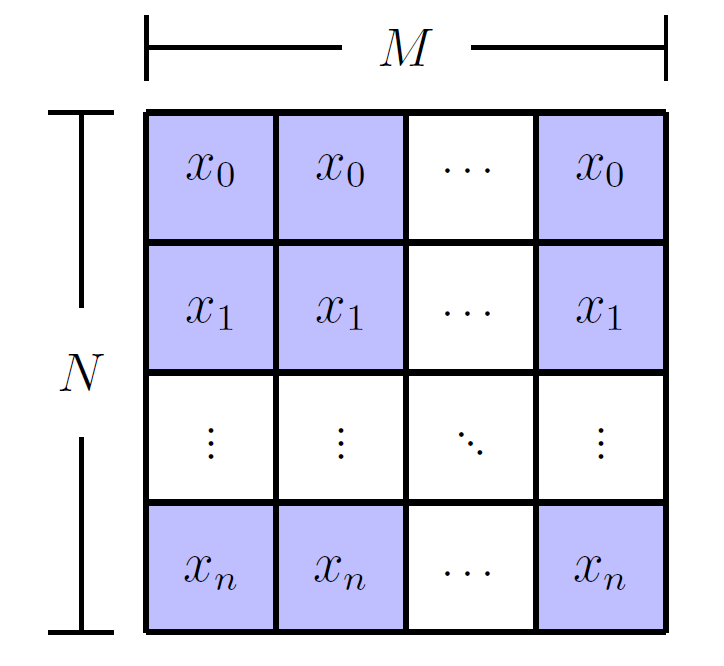}
\caption{Interleaved data format of the array containing the RHS matrix $\mathbf{b}$ used in this paper.}
\label{fig:interleaved_data}
\end{figure}

\section{Matrix Solver Methodology}
\label{method}
The methodologies implementing the tridiagonal and pentadiagonal algorithms we are concerned with in this paper have to date relied on using interleaved data layouts 
(shown schematically in Figure~\ref{fig:interleaved_data}) in global memory.
Each thread handles a particular system solution and the system is solved by a forward sweep followed by a backward sweep.  
The interleaved data format is to ensure coalescence of data accesses during theses sweeps, with each thread retrieving neighbouring memory locations in global memory for various $\mxA$ and $\vecb$ matrix entries~\cite{cuThomasBatch}.
This maximises bandwidth but is limiting to the overall memory budget as each system requires its own copy of the $\mxA$ matrix.
Instead we propose to store globally just a single copy of the $\mxA$ matrix entries with all threads accessing the desired value at the same time, the right-hand-side (RHS) matrices, $\vecb$, will still be stored in the same interleaved data access pattern as before.
This will reduce the overall bandwidth of memory accesses to the $\mxA$ but this will not harm the solver's performance as we will show in later sections.
Critically also this approach of using just a single $\mxA$ matrix will save drastically on the amount of memory used in batch solvers allowing for larger batch numbers and greater matrix sizes to be solved on a single GPU, increasing hardware usage efficiency.

We now discuss the memory access pattern for the single $\mxA$ matrix.
Global memory access in GPUs is achieved via the L2 cache.
When a piece of memory is requested for by a warp this memory is retrieved from global memory and copied to the L2 cache followed by the L1 cache on the SM before then being used by the relevant warps being computed on that SM.
Each SM has their own pipe to the L2 cache. 
So when warps from different SMs all request the same memory they will all get a cache hit in at least the L2 cache except the first one to arrive which will be the warp to retrieve the data from the global memory on the device.
We can guarantee that this will occur as each warp will be computing the same instruction in either the forward or backward sweep of the solver but warps on different SMs will not arrive to the L2 cache at the same time.
In addition to this warps which share L1 caches will get cache hits here when they're not the first to request the piece of memory. 
A diagram of this memory access pattern discussion can also be seen in Figure~\ref{memoryaccess}.
Thus it is clear that given every piece of memory must follow the above path most of the threads solving a batched problem will benefit from speed-ups due to cache hits as they no longer need to retrieve their own copy of the data.

In the following two sections we present specific details for each of the tridiagonal and pentadiagonal schemes along with the previously discussed benchmarks.
The benchmarks are performed on 16GB Tesla V100 GPUs with the pre-factorisation step for the single $\mxA$ performed on the CPU, this pre-factorisation step is a purely serial operation and hence the CPU was chosen to perform this computation.
We shall refer to the new versions of the algorithms as cuThomasConstantBatch and cuPentConstantBatch for tridiagonal and pentadiagonal respectively.
We use the nomenclature `Constant' denoting the fact that all systems have the same $\mxA$. 
The $\vecb$ will be stored as usual in an interleaved data format and is computed using cuSten~\cite{gloster2019custen}. In both algorithms we utilise uniform memory access. In terms of thread organisation structure, two grids were used. The first, a 1D grid, performs the inversion (Thomas algorithm). The second is a 2D gird which performs the updating steps of both algorithms.
For notational purposes we will use $N$ to describe the number of unknowns in our systems and $M$ to describe the batch size (number of systems being solved).


\section{Tridiagonal Systems}
\label{tri}
In this section we first present the Thomas Algorithm~\cite{numalgC} and how it is modified for batch solutions with multiple $\vecb$ matrices and with a single $\mxA$. 
We then present our chosen benchmark problem of periodic diffusion equations and then the results.

\subsection{Tridiagonal Inversion Algorithm}
We begin with a generalised tridiagonal matrix system $\mxA \vecx=\vecrhsT$ with $\mxA$ given by
\begin{equation*}
\mxA = 
\begin{pmatrix}
b_1 & c_1 & 0 & \cdots &   \cdots & 0 \\
a_2 & b_2 & c_2 & 0 & \cdots   & \vdots \\
0 & a_3 & b_3 & c_3 & 0 &  \vdots \\
\vdots & \ddots & \ddots & \ddots & \ddots & 0  \\
\vdots & \cdots  & 0   & a_{N - 1}  & b_{N - 1} & c_{N - 1} \\
0 & \cdots  &   \cdots & 0 & a_{N}  & b_{N} 
\end{pmatrix}.
\end{equation*}
We then solve this system using a pre-factorisation step followed by a forwards and backwards sweep. 
The pre-factorisation is given by
\begin{equation}
\hat{c}_1 = \frac{c_1}{b_1}
\end{equation}
And then for $i = 2, \dots, N$
\begin{equation}
\hat{b}_i = b_i - a_i \hat{c}_{i - 1} \qquad \hat{c}_i = \frac{c_i}{\hat{b}_i}
\end{equation}
For the forwards sweep we have
\begin{equation}
\hat{d}_1 = \frac{d_1}{b_1}
\end{equation}
\begin{equation}
\hat{d}_i = \frac{d_i - a_i \hat{d}_{i - 1}}{\hat{b}_i}
\end{equation}
While for the backwards sweep we have 
\begin{equation}
x_N = \hat{d}_N
\end{equation}
And for $i = N-1, \dots, 1$
\begin{equation}
x_i = \hat{d}_i - a_i \hat{c}_i
\end{equation}
A summary of cuThomasConstantBatch can be found in Appendix A. The pre-factorization step performed is performed on the CPU and the remaining forward and backward sweeps of the algorithm which are performed on the GPU can be found in Algorithm 1 and Algorithm 2 in Appendix A.

Previous applications of this algorithm~\cite{cuThomasBatch} required that each thread had access to its own copy of $4$ vectors, the $3$ diagonals $a_i$, $b_i$ and $c_i$ along with the RHS $d_i$.
These would then be overwritten in the pre-factorisation and solve steps to save memory, thus the total memory usage here is $O(4 \times M \times N)$.
We now limit the $\mxA$ to a single global case that will be accessed simultaneously using all threads as discussed in Section~\ref{method} and retain the individual $\vecb$ in interleaved format for each thread $f_i$.
This reduces the data storage to $O(3 \times N + M \times N)$, an approximate 75\% reduction.
We present benchmark methodology and results for this method in the following subsections.

\subsection{Benchmark Problem}
For a benchmark problem we solve the Diffusion Equation, a standard model equation in Computational Science and Engineering, its presence can be seen in most systems involving heat and mass transfer.
The equation on a one-dimensional interval $(0,L)$ is given as:
\begin{equation}
\frac{\partial C}{\partial t} = \alpha \frac{\partial^2 C}{\partial x^2},\qquad t>0,\qquad x\in (0,L),
\label{eq:diffusion}
\end{equation} 
where $\alpha$ is the diffusion coefficient.
We solve this equation on a periodic domain of length $L$ such that $C(x + L,t) = C(x,t)$ with an initial condition $C(x, t = 0) = f(x)$ valid on the domain $[0,L]$.
We rescale Equation~\eqref{eq:diffusion} by setting $\alpha = 1$ and $L = 1$ and integrate the equation forward in time using a standard Crank--Nicolson scheme with central differences for space which is unconditionally stable.
Finite differencing is done using standard notation
\begin{equation}
C_i^n = C(x=i\Delta x ,t=n\Delta t)
\label{eq:stdDiff}
\end{equation}
where $\Delta x = L / N$ and $i = 1 \dots N$.
Thus our numerical scheme can be written as 
\begin{equation}
- \sigma_x C_{i - 1}^{n+1} + (1 + 2 \sigma_x) C_{i}^{n+1} - \sigma_x C_{i+1}^{n+1}
=  \sigma_x C_{i - 1}^{n} + (1 - 2 \sigma_x)C_{i}^{n} + \sigma_x C_{i+1}^{n} 
\label{eq:1ddiffscheme}
\end{equation} 
where
\begin{equation}
\sigma_x = \frac{\Delta t}{2 \Delta x^2}
\end{equation}
Thus we can relate these coefficients to matrix entries by
\begin{equation}
a_i  = - \sigma_x,\qquad b_i = 1 + 2 \sigma_x,\qquad
c_i = - \sigma_x
\end{equation}%
In the next section we present our method to deal with the periodicity of the matrix and then after that present the benchmark comparison with the existing state of the art.

\subsection{Periodic Tridiagonal Matrix}
As the system is periodic two extra entries will appear in the matrix, one in the top right corner and another in the bottom left, thus our matrix is now given by
\begin{equation*}
\mxA = 
\begin{pmatrix}
b & c & 0 & \cdots &  \cdots & 0 & a \\
a & b & c & 0 & \cdots & \cdots   & 0 \\
0 & a & b & c & 0 & \cdots  & \vdots \\
\vdots & \ddots & \ddots & \ddots & \ddots & \ddots &   \vdots\\
\vdots & \cdots  & 0  &  a  & b  & c & 0  \\
0 & \cdots & \cdots  & 0   & a  & b & c \\
c & 0 & \cdots  &   \cdots & 0 & a  & b 
\end{pmatrix}.
\end{equation*}
To deal with these we use the Sherman--Morrison formula: we rewrite our system as 
\begin{equation}
\mxA\vecx  = (\mxA' + \mxU \otimes \mxV) \vecx = \vecrhsT. 
\end{equation} 
Here,
\begin{subequations}
\begin{equation}
\mxU = 
\begin{pmatrix}
-b \\
0 \\
\vdots \\
\vdots \\
0 \\
c\\
\end{pmatrix},
\qquad
\mxV = 
\begin{pmatrix}
1 \\
0 \\
\vdots \\
\vdots \\
0 \\
- a / b\\
\end{pmatrix},
\qquad
\end{equation}
and
\begin{equation}
\mxA' = 
\begin{pmatrix}
2b & c & 0 & \cdots &  \cdots & 0 & 0 \\
a & b & c & 0 & \cdots & \cdots   & 0 \\
0 & a & b & c & 0 & \cdots  & \vdots \\
\vdots & \ddots & \ddots & \ddots & \ddots & \ddots &   \vdots\\
\vdots & \cdots  & 0  &  a  & b  & c & 0  \\
0 & \cdots & \cdots  & 0   & a  & b & c \\
0 & 0 & \cdots  &   \cdots & 0 & a  & b + a c / b
\end{pmatrix}.
\end{equation}%

\end{subequations}%
Thus two tridiagonal systems must now be solved
\begin{equation}
\mxA' \vecy = \vecrhsT  \qquad \mxA' \vecz = \mxU
\end{equation}
the second of which need only be performed once at the beginning of a given simulation.
Finally to recover $\vecx$ we substitute these results into 

\begin{equation}
\vecx = \vecy - \left(\frac{\mxV \cdot \vecy}{1 + (\mxV \cdot \vecz)}\right)\vecz
\end{equation}

\begin{figure}[htb]
\centering
\includegraphics[width=0.7\textwidth]{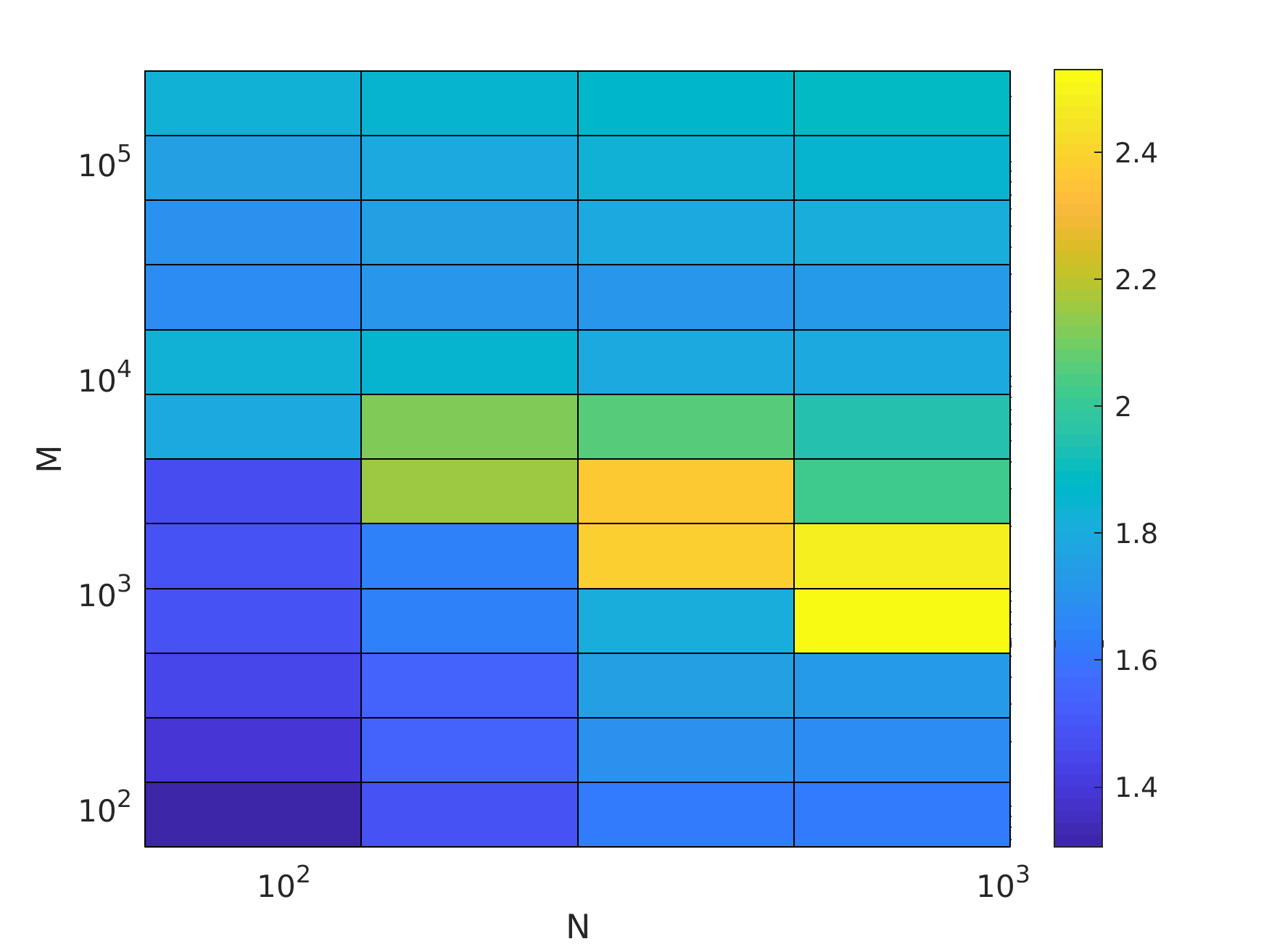}
\caption{Speedup of of cuThomasConstant versus cuThomasBatch (gtsvInterleavedBatch) }
\label{speedupInterleaved}
\end{figure}

\begin{figure}[htb]
\centering
\includegraphics[width=0.7\textwidth]{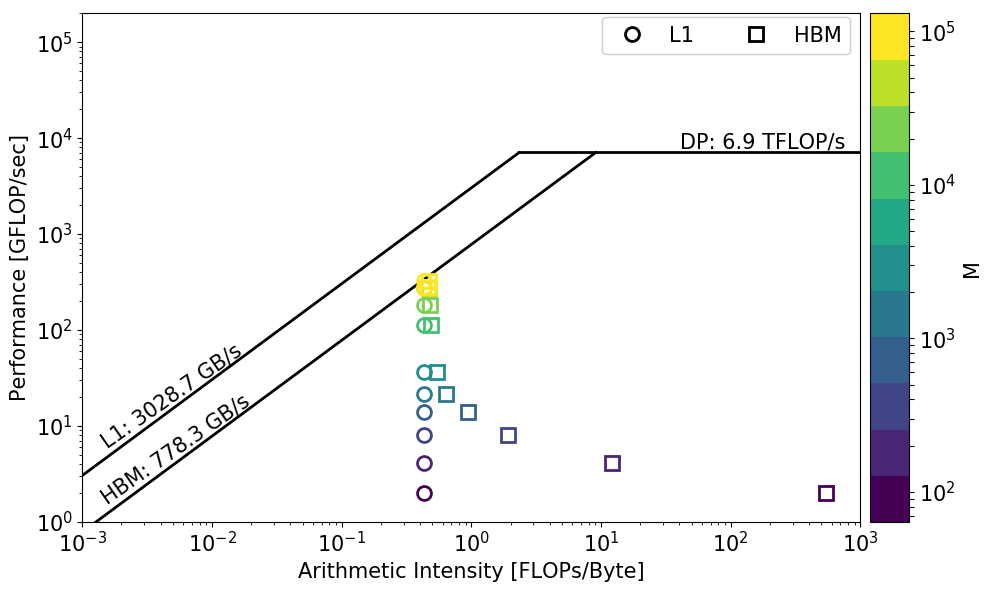}
\caption{Roofline plot of the cuThomasConstant kernel for $N = 1024$ }
\label{cuThomasConstantRoofline}
\end{figure}

\subsection{Benchmark Results}

The diffusion equation benchmark problem presented above is solved for 1000 time-steps in order to extract the relevant timing statistics and average out any effects of background processes on the benchmark.
We omit start up costs such as memory allocation, setting of parameters etc. We also do not include the timing of the pre-factorization step. This operation is only performed once and its cost is so insignificant to the rest of the algorithm that we attribute it to part of the setup cost. As such, both codes are timed for just the time--stepping sections of the process. The speed-up results can be seen in Figure~\ref{speedupInterleaved}, in all situations there is a clear speed-up over the existing state of the art cuThomasBatch. 
The largest speed-ups are for $N$ large and moderate $M$.
Significant speed-up is available for all of large $M$ which is the domain where GPUs would typically be deployed to solve the problem.  Shown also in Figure~\ref{cuThomasConstantRoofline} is a roofline plot of the cuThomasConstantBatch kernel: it is evident that as the batch size is increased the kernel tends towards the  memory roofline of the device.

It should be noted that some of the speed-up seen here can be attributed to the pre-factorisation step which the cuThomasBatch implementation does not have. 
cuThomasBatch requires that the $\mxA$ matrix be reset at every time-step as the pre-factorisation and solve steps are carried out in the one function, leading to an overwrite of the data and making it unusable for repeated time-stepping.
This overwriting feature has been seen in previous studies~\cite{gloster2019cupentbatch} where the authors also carried out a rewrite to make the benchmarking conditions between their work and the state of the art.
It was shown that not all of the speed-up can be attributed to the lack of needing to reset the $\mxA$ matrix, thus some of the performance increase we are seeing in Figure~\ref{speedupInterleaved} can be attributed to the new data layout presented in this paper.
Our second benchmark for the pentadiagonal case will also show this as there was no resetting of the matrix required.

\begin{table}[]
\caption{Table of execution times for cuThomasConstantBatch and cuThomasBatch (measured in seconds).}
\centering
\setlength{\tabcolsep}{10pt}
\begin{small}
\begin{tabular}{cc|c|c}
\hline
\multicolumn{1}{c}{}                                                 & \multicolumn{1}{c|}{}           & \multirow{2}{*}{\textbf{cuThomasConstantBatch}} & \multirow{2}{*}{\textbf{cuThomasBatch}} \\
\multicolumn{1}{c}{\textbf{M}}                                       & \multicolumn{1}{c|}{\textbf{N}} &                                                 &                                         \\ \hline
\hline
\multicolumn{1}{l|}{\multirow{3}{*}{\textbf{$2^6$}}}                 & 64                              & 0.097                                           & 0.127                                   \\
\multicolumn{1}{l|}{}                                                & 256                             & 0.161                                           & 0.267                                   \\
\multicolumn{1}{l|}{}                                                & 1024                            & 0.510                                           & 0.850                                   \\ \hline
\multicolumn{1}{l|}{\multirow{3}{*}{\textbf{$2^{12}$}}}              & 64                              & 0.101                                           & 0.192                                   \\
\multicolumn{1}{l|}{}                                                & 256                             & 0.255                                           & 0.549                                   \\
\multicolumn{1}{l|}{}                                                & 1024                            & 1.020                                           & 1.987                                   \\ \hline
\multicolumn{1}{l|}{\multirow{3}{*}{\textbf{$2^{16}$}}}              & 64                              & 1.278                                           & 2.894                                   \\
\multicolumn{1}{l|}{}                                                & 256                             & 5.015                                           & 11.460                                  \\
\multicolumn{1}{l|}{}                                                & 1024                            & 20.268                                          & 46.175                                  \\ \hline
\hline
\end{tabular}
\end{small}
\label{table:diff_times}
\end{table}


\section{Pentadiagonal Systems}
\label{pent}
In this section we first present a modified version of the pentadiagonal inversion algorithm as presented in~\cite{gloster2019cupentbatch} with a single $\mxA$. 
Then we present the benchmark problem and finally this is followed by the results comparing our new implementation with existing state of the art.

\subsection{Pentadiagonal Inversion Algorithm}
In this section we describe a standard numerical method~\cite{numalgC, gloster2019cupentbatch} for solving a pentadiagonal problem $\mxA\vecx=\vecrhs$.  We present the algorithm in a general context so we have a pentadiagonal matrix given by

\begin{equation*}
\mxA = 
\begin{pmatrix}
c_1 & d_1 & e_1 & 0 & \cdots &  0 & \cdots & 0 \\
b_2 & c_2 & d_2 & e_2 & 0 & \cdots & \cdots   & \vdots \\
a_3 & b_3 & c_3 & d_3 & e_3 & 0 & \cdots  & 0\\
0 & \ddots & \ddots & \ddots & \ddots & \ddots & \ddots &   \vdots\\
\vdots& \ddots & \ddots & \ddots & \ddots & \ddots & \ddots & 0 \\
0 & \cdots  & 0  &  a_{N - 2}  & b_{N - 2}  & c_{N - 2} & d_{N - 2} & e_{N - 2} \\
0 & \cdots & \cdots  & 0   & a_{N - 1}  & b_{N - 1} & c_{N - 1} & d _{N - 1}\\
0 & \cdots & \cdots  &   \cdots & 0 &  a_{N} & b_{N}  & c_{N} 
\end{pmatrix}.
\end{equation*}
Three steps are required to solve the system:
\begin{enumerate}
\item Factor $\mxA = \mxL\mxR$  to obtain $\mxL$ and $\mxR$.
\item Find $\vecg$ from $\vecf = \mxL\vecg$
\item Back-substitute to find $\vecx$ from $\mxR\vecx = \vecg$
\end{enumerate}
Here, $\mxL$, $\mxR$ and $\vecg$ are given by the following equations:
\begin{subequations}
\begin{equation}
\mxL = 
\begin{pmatrix}
\alpha_1 &  &  &  &  &   &   \\
\beta_2 & \alpha_2 &  &  &  &  &     \\
\epsilon_3 & \beta_3 & \alpha_3 &  &  &  &   \\
 & \ddots & \ddots & \ddots &  &  &     \\
 &  &  \epsilon_{N - 1}  & \beta_{N - 1}  & \alpha_{N - 2} &    \\
&  &    & \epsilon_{N - 1}  & \beta_{N - 1} & \alpha_{N }\\
\end{pmatrix},
\qquad
\vecg = \begin{pmatrix}
g_{1} \\
g_{2} \\
\vdots \\
\vdots \\
g_{N-1} \\
g_{N}
\end{pmatrix},
\end{equation}
\begin{equation}
\mxR = 
\begin{pmatrix}
1 & \gamma_1  & \delta_1  &  &  &   &   \\
 & 1 & \gamma_2  & \delta_2  &  &  &     \\
 &  & \ddots & \ddots & \ddots  &  &   \\
 & & & 1  & \gamma_{N-2}  & \delta_{N-2}    \\
 &  &  &  & 1 & \gamma_{N-1}    \\
&  &    & &  & 1\\
\end{pmatrix}
\end{equation}%
\end{subequations}%
(the other entries in $\mxL$ and $\mxR$ are zero).  
The explicit factorisation steps for the factorisation $\mxA=\mxL\mxR$ are as follows:
\begin{enumerate}
\item $\alpha_1 = c_1$
\item $\gamma_1 = \frac{d_1}{\alpha_1}$
\item $\delta_1 = \frac{e_1}{\alpha_1}$
\item $\beta_2 = b_2$
\item $\alpha_2 = c_2 - \beta_2\gamma_1$
\item $\gamma_2 = \frac{d_2 - \beta_2 \delta_1}{\alpha_2}$
\item $\delta_2 = \frac{e_2}{\alpha_2}$
\item For each $i = 3, \dots, N-2$
\begin{enumerate}
\item $\beta_i = b_i - a_i \gamma_{i-2}$
\item $\alpha_i = c_i - a_i\delta_{i-2} - \beta_i \gamma_{i-1}$
\item $\gamma_i = \frac{d_i - \beta_i \delta_{i-1}}{\alpha_i}$
\item $\delta_i = \frac{e_i}{\alpha_i}$
\end{enumerate}
\item $\beta_{N-1} = b_{N-1} - a_{N - 1}\gamma_{N-3}$
\item $\alpha_{N - 1} =  c_{N-1} - a_{N-1}\delta_{N-3} - \beta_{N-1}\gamma_{N-2}$
\item $\gamma_{N-1} = \frac{d_{N-1}-\beta_{N-1}\delta_{N-2}}{\alpha_{N-1}}$
\item $\beta_{N} = b_{N} - a_{N }\gamma_{N-2}$
\item $\alpha_{N} =  c_{N}- a_{N}\delta_{N-2} - \beta_{N}\gamma_{N-1}$
\item $\epsilon_i = a_i, \quad \forall i$
\end{enumerate}
The steps to find $\vecg$ are as follows:
\begin{enumerate}
\item $g_1 = \frac{f_1}{ \alpha_1}$
\item $g_2 = \frac{f_2 - \beta_2 g_1}{\alpha_2}$
\item  $g_i = \frac{f_i - \epsilon_i g_{i-2} - \beta_i g_{i - 1}}{\alpha_i} \quad \forall i = 3 \cdots N$
\end{enumerate}
Finally, the back-substitution steps  find $\vecx$ are as follows:
\begin{enumerate}
\item $x_N = g_N$
\item $x_{N-1} = g_{N-1} - \gamma_{N-1}x_N$
\item  $x_i = g_i - \gamma_i x_{i+1} - \delta_{i}x_{i+2} \quad \forall i = (N-2) \cdots 1$
\end{enumerate}

Previous applications of this algorithm~\cite{gloster2019cupentbatch} required that each thread had access to its own copy of $6$ vectors, the $5$ diagonals $a_i$, $b_i$, $c_i$, $d_i$ and $e_i$ along with the $\vecb$ $f_i$.
These would then be overwritten in the pre-factorisation and solve steps to save memory, thus the total memory usage here is $O(6 \times M \times N)$.
We now limit the $\mxA$ to a single global case that will be accessed simultaneously using all threads as discussed in Section~\ref{method} and retain the individual $\vecb$ in interleaved format for each thread $f_i$.
This reduces the data storage to $O(5 \times N + M \times N)$, this is an approximate 83\% reduction in data usage.
We present benchmark methodology and results for this method in the following subsections.

\subsection{Benchmark Problem}

\begin{figure}[htb]
\centering
\includegraphics[width=0.7\textwidth]{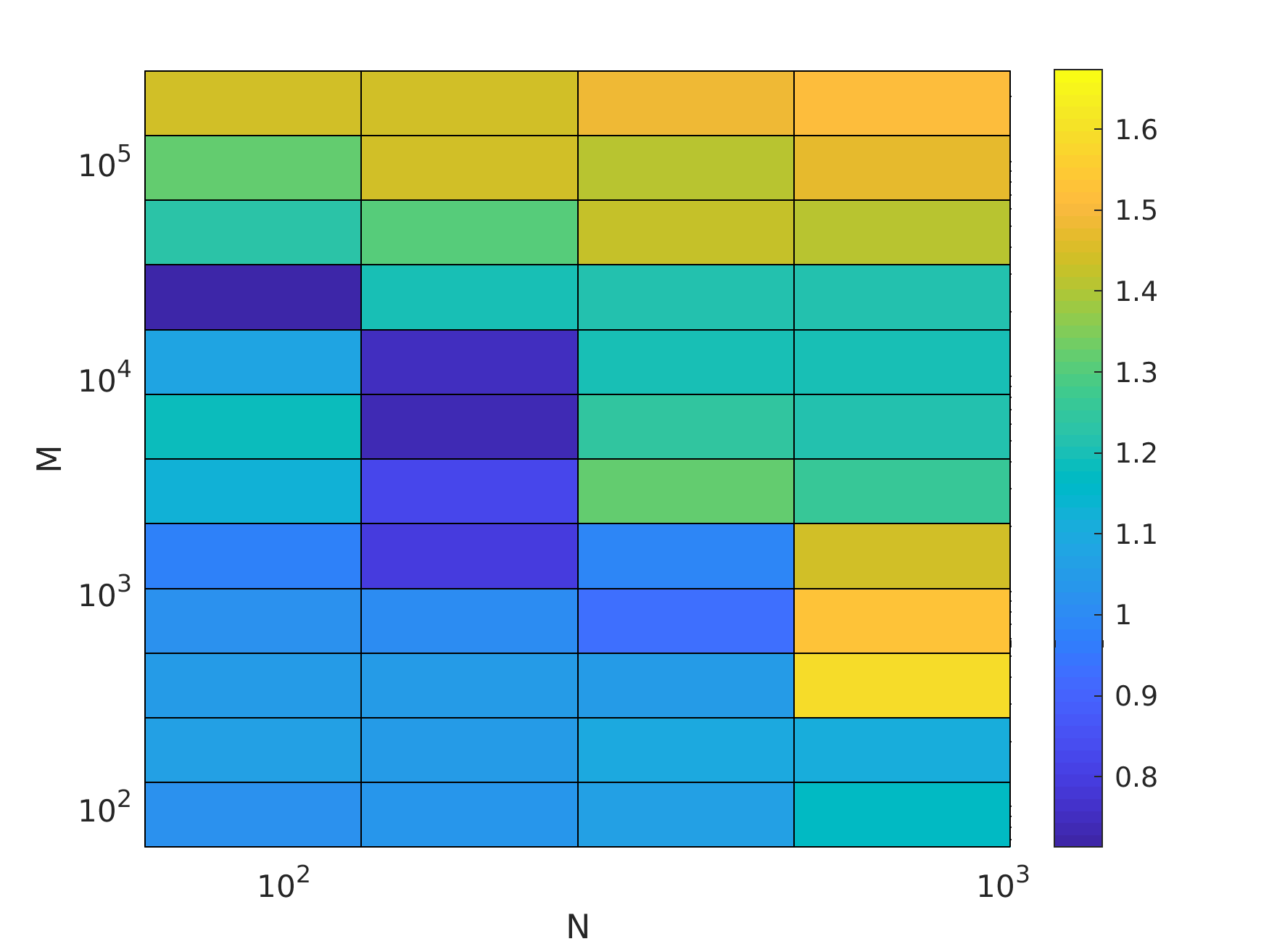} 
\caption{Speedup of of cuPentConstantBatch versus cuPentBatch}
\label{pentConstantSpeedUp}
\end{figure}

\begin{figure}[htb]
\centering
\includegraphics[width=0.7\textwidth]{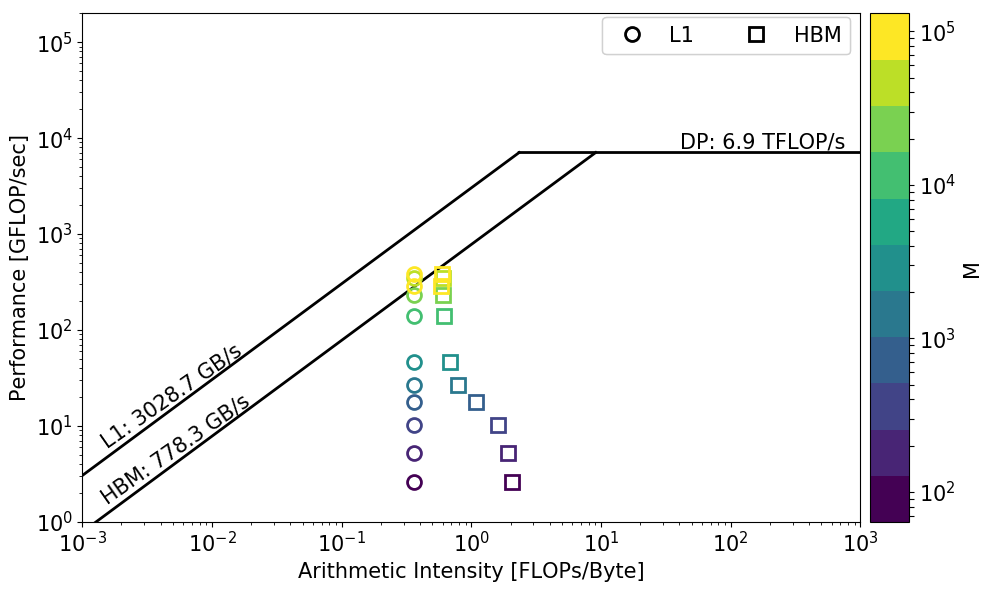} 
\caption{Roofline plot of the cuPentConstantBatch kernel for $N = 1024$}
\label{pentConstantRoofline}
\end{figure}

For a benchmark problem we solve the hyperdiffusion equation in a one-dimensional interval $(0,L)$, given here as:
\begin{equation}
\frac{\partial C}{\partial t} = -\alpha \frac{\partial^4 C}{\partial x^4},\qquad t>0,\qquad x\in(0,L),
\label{eq:hyperdiffusion}
\end{equation} 
where $\alpha$ is the hyperdiffusion coefficient.
We solve this equation on a periodic domain of length $L$ such that $C(x + L,t) = C(x,t)$ with an initial condition $C(x, t = 0) = f(x)$ valid on the domain $x\in [0,L]$.
We rescale Equation~\eqref{eq:hyperdiffusion} by setting $\alpha = 1$ and $L = 1$ and integrate the equation forward in time using a standard Crank--Nicolson scheme with central differences for space which is unconditionally stable.
Finite differencing is done the same standard notation as in Section~\ref{tri} (specifically, Equation~\eqref{eq:stdDiff}).  
Thus, our numerical scheme can be written as 
\begin{multline}
\sigma_x C_{i - 2}^{n + 1} - 4 \sigma_x C_{i - 1}^{n+1} + (1 + 6 \sigma_x)C_{i}^{n+1} - 4 \sigma_x C_{i+1}^{n+1} + \sigma_x C_{i+2}^{n+1} \\ 
=  - \sigma_x C_{i - 2}^{n} + 4 \sigma_x C_{i - 1}^{n} + (1 - 6 \sigma_x)C_{i}^{n} + 4 \sigma_x C_{i+1}^{n} - \sigma_x C_{i+2}^{n},
\label{eq:1d_hyper_scheme}
\end{multline} 
where now $\sigma_x$ means
\begin{equation}
\sigma_x = \frac{\Delta t}{2 \Delta x^4}
\end{equation}
For a more more detailed exposition of this benchmark problem, see Reference~\cite{gloster2019cupentbatch}.


\subsection{Benchmark Results}

We plot the speed-up of cuPentConstantBatch versus cuPentBatch solving batches of the above method for the hyperdiffusion equation in Figure~\ref{pentConstantSpeedUp}.  
Like cuThomasConstantBatch, the pre-factorization step is not included in these timings. We use cuSten~\cite{gloster2019custen}  to calculate the values for $\vecrhs$. 
We also present a roofline plot for the the cuPentConstantBatch kernel in Figure~\ref{pentConstantRoofline}.  Similar to cuThomasConstantBatch, the cuPentConstantBatch kernel is ultimately limited by the memory constraint of the device.

From Figure~\ref{pentConstantSpeedUp}, it can be seen that cuPentBatchConstant outperforms cuPentBatch consistently for high values of both $M$ and $N$. 
At low $M$ and $N$ the two methods are roughly equivalent; indeed, there are some areas where cuPentBatch performs better.  However, in this part of the parameter space,  standard CPU implementation is more sensible -- taking into account the movement over the PCI lane along with other overheads.
Furthermore, if we examine much larger values of $M$ and $N$ than those plotted, the superior performance of cuPentBatchConstant becomes much more apparent.  The performance difference in these cases can be orders of magnitude, as the memory for cuPentBatch rapidly exceeds the available memory on the GPU while cuPentBatchConstant can still fit.  We explore this part of the parameter space much more thoroughly in Section~\ref{1D}.
%


\begin{table}[]
\centering
\caption{Table of execution times for cuPentConstantBatch and cuPentBatch (measured in seconds).}
\setlength{\tabcolsep}{10pt}
\begin{small}
\begin{tabular}{cc|c|c}
\hline
\multicolumn{1}{c}{}                                                 & \multicolumn{1}{c|}{}           & \multirow{2}{*}{\textbf{cuPentConstantBatch}} & \multirow{2}{*}{\textbf{cuPentBatch}}   \\
\multicolumn{1}{c}{\textbf{M}}                                       & \multicolumn{1}{c|}{\textbf{N}} &                                                 &                                           \\ \hline
\hline
\multicolumn{1}{l|}{\multirow{3}{*}{\textbf{$2^6$}}}                 & 64                              & 369.902                                         & 376.321                                   \\
\multicolumn{1}{l|}{}                                                & 256                             & 461.727                                         & 487.455                                   \\
\multicolumn{1}{l|}{}                                                & 1024                            & 658.475                                         & 717.109                                   \\ \hline
\multicolumn{1}{l|}{\multirow{3}{*}{\textbf{$2^{12}$}}}              & 64                              & 460.857                                         & 523.693                                   \\
\multicolumn{1}{l|}{}                                                & 256                             & 429.127                                         & 535.657                                   \\
\multicolumn{1}{l|}{}                                                & 1024                            & 1344.377                                        & 1495.057                                  \\ \hline
\multicolumn{1}{l|}{\multirow{3}{*}{\textbf{$2^{16}$}}}              & 64                              & 1786.018                                        & 2562.037                                  \\
\multicolumn{1}{l|}{}                                                & 256                             & 6535.750                                        & 9707.131                                  \\
\multicolumn{1}{l|}{}                                                & 1024                            & 25893.904                                       & 38407.516                                 \\ \hline
\hline
\end{tabular}
\end{small}
\label{table:hyper_times}
\end{table}


\section{Physical Application -- Coarsening and the Cahn--Hilliard Equation}
\label{1D}

In this section, we study the Cahn--Hilliard equation in one spatial dimension.  The purpose of this section is twofold -- we wish to showcase the GPU methodology as applied to a meaningful physical problem.  We also wish to explore the coarsening phenomenon of the one-dimensional Cahn--Hilliard equation, in a statistically robust fashion.  This can be readily achieved with the present GPU methodology, which enables the simulation of $2^{20}$ individual copies of the Cahn--Hilliard equation.

We begin by presenting the Cahn--Hilliard equation in one spatial dimension, valid on an interval $(0,L)$:
\begin{equation}
\frac{\partial C}{\partial t} = \frac{\partial^2}{\partial x^2} \left(C^3 - C - \gamma \frac{\partial^2 C}{\partial x^2} \right),\qquad t>0,\qquad x\in (0,L).
\label{eq:1Dcahn}
\end{equation}
Here, $C$ is a scalar concentration field taking arbitrary real values.  The concentration field is used to demarcate different domains of a binary mixture~\cite{CH_orig}.  Thus, a local solution $C\approx 1$ indicates a region rich in one binary fluid component, whereas a local solution $C\approx -1$ indicates a region rich int he other fluid component.  Here also, $\gamma$ is a positive constant; the domains are separated by transition regions of width $\sqrt{\gamma}$.  We assume for definiteness that Equation~\eqref{eq:1Dcahn} is posed on the interval $(0,L)$ with periodic boundary conditions, although other standard boundary conditions (e.g.  Neumann, Dirichlet) are possible also.

With such standard boundary conditions, the mean value of the concentration is conserved:
\begin{equation}
\frac{\mathd}{\mathd t}\langle C\rangle=0,\qquad \langle C\rangle=\frac{1}{L}\int_0^L C(x,t)\,\mathd x.
\label{eq:conserved}
\end{equation}
This can be seen by doing differentiation under the integral in Equation~\eqref{eq:conserved} and replacing $\partial C/\partial t$ with $\partial_x^2\left(C^3-C-\gamma\partial_x^2C\right)$.  Application of the periodic boundary conditions to the resulting integral gives 
\begin{equation}
\frac{\mathd}{\mathd t}\langle C\rangle=\frac{1}{L}\int_0^L  \partial_x^2\left(C^3-C-\gamma\partial_x^2C\right)\mathd x=0,
\end{equation}
hence $\langle C\rangle=\mathrm{Const.}$.  The value of $\langle C\rangle$ is a a key parameter in this section and in Section~\ref{flowpattern}.

\begin{figure}[htb]
\centering
\includegraphics[width=0.8\textwidth]{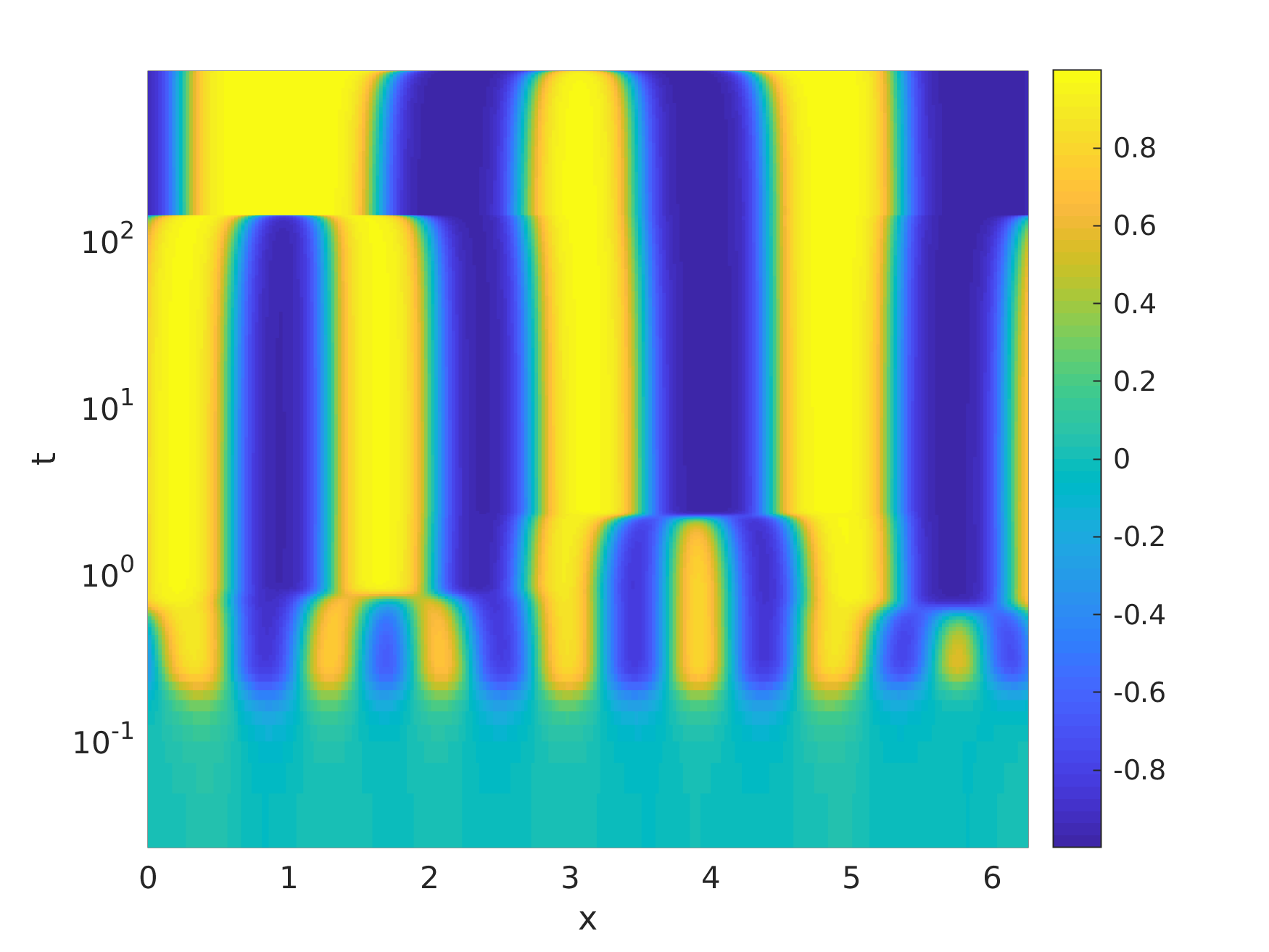}
\caption{
Space time plot showing the time evolution of the solution of the 1D Cahn--Hilliard equation.  Simulation parameters: $\gamma = 0.01$, domain size $L=2\pi$, $N=256$ grid-points, such that $\Delta x=L/N\ll \sqrt{\gamma}$, to properly resolve the transition zones between domains.  
The initial condition is $C(x,t=0)=0+\text{Random fluctuations}$, the random fluctuations are numbers drawn from the uniform distribution on the interval $[-0.1,0.1]$.
The simulation is run out to a final time $T=1000$ and the numerical scheme is the one presented in Equation~\eqref{eq:1Dnumerical}.}
\label{1Dspacetime}
\end{figure}

Any constant value $C(x,t)=C_0$ is a solution of Equation~\eqref{eq:1Dcahn}.  This constant solution can be characterized by introducing a perturbed solution
\begin{equation}
C(x,t=0)=C_0+\delta(x),
\label{eq:perturb}
\end{equation}
where $\delta(x)$ is a mean-zero perturbation and is `small' in the sense that $\max(|\delta|^2)\ll \max(|\delta|)$.  As such,
the solution~\eqref{eq:perturb} with $C_0^2>1/3$ is a stable solution to Equation~\eqref{eq:1Dcahn}, in the sense of linear stability analysis; that is, $\lim_{t\rightarrow \infty}C(x,t)=C_0$.  In contrast, the solution~\eqref{eq:perturb} with $C_0^2<1/3$ is unstable, and furthermore has the property that $\lim_{t\rightarrow \infty}C(x,t)=\pm 1$, in domains.  The domains correspond to phase separation -- regions rich in one fluid component or the other, and are separated by transition zones of width $\sqrt{\gamma}$.  The spontaneous formation of domains from the initial condition~\eqref{eq:perturb} in this instance is called \textit{coarsening}.  A sample numerical simulation showing the coarsening phenomenon is shown in Figure~\ref{1Dspacetime}.


The coarsening phenomenon in the Cahn--Hilliard equation is dimension-dependent.  In one dimension, theoretical arguments~\cite{argentina2005coarsening} predict that the size of a typical domain $\typical(t)$ should grow as $\typical(t)\sim \ln (t)$, at late times.  In contrast, in higher dimensions, the size of a typical domain grows as $\typical(t)\sim t^{1/3}$, which is the famous Lifshitz--Slyozov coarsening law~\cite{LS}.  The growth of the `typical' domain size in this way is referred to as \textit{scaling}.  As such, the aim of this section is to confirm the theoretical one-dimensional scaling law.

\subsection{1D Numerical Scheme}
\label{1Dscheme}

Equation~\eqref{eq:1Dcahn} is solved implicitly on a uniform grid.  We  use a superscript $n$ to denote evaluation of the solution at time $t=n\Delta t$.
Taking the hyperdiffusion term to the $\mxA$, and discretising in time we get
\begin{equation}
C^{n+1} + \Delta t \gamma \frac{\partial^4 C^{n+1}}{\partial x^4} = C^n + \Delta t \left(\frac{\partial^2}{\partial x^2} \left(C^3 - C \right)^n \right)
\label{eq:1Dnumerical}
\end{equation}
%
We apply standard second-order accurate differencing to the spatial derivatives:
\begin{subequations}
\begin{equation}
\frac{\partial^2 C}{\partial x^2} \approx \delta_x^2 C = \frac{C_{i + 1} - 2 C_i + C_{i-1}}{\Delta x^2}
\end{equation}
\begin{equation}
\frac{\partial^4 C}{\partial x^4} \approx \delta_x^4 C = \frac{C_{i - 2} - 4 C_{i-1} + C_i - 4 C_{i+1} + C_{i + 2}}{\Delta x^4}
\end{equation}
\end{subequations}
This, along with the periodicity of the domain, yields a pentadiagonal matrix system of the form
\begin{subequations}
\begin{equation}
\underbrace{
\begin{pmatrix}
c & d & e & 0 & \cdots &  0 & a & b \\
b & c & d & e & 0 & \cdots &   0 & a \\
a & b & c & d & e & 0 & \cdots  & 0\\
0 & \ddots & \ddots & \ddots & \ddots & \ddots & \ddots &   \vdots\\
\vdots& \ddots & \ddots & \ddots & \ddots & \ddots & \ddots & 0 \\
0 & \cdots  & 0  &  a  & b  & c & d & e \\
e & 0 &  & 0   & a  & b & c & d \\
d & e & 0  &   \cdots & 0 &  a & b & c
\end{pmatrix}
}_{=\mxA}
\underbrace{
\begin{pmatrix}
C_{1} \\
C_{2} \\
\vdots \\
\vdots \\
\vdots \\
C_{N-2} \\
C_{N-1} \\
C_{N}
\end{pmatrix}
}_{=\vecx}
=
\underbrace{
\begin{pmatrix}
f_{1} \\
f_{2} \\
\vdots \\
\vdots \\
\vdots \\
f_{N-2} \\
f_{N-1} \\
f_{N}
\end{pmatrix}
}_{=\vecrhs}.
\label{eq:matrix_sys1}
\end{equation}
Here, the coefficients of the matrix in Equation~\eqref{eq:matrix_sys1} have the following meaning:
\begin{equation}
a  = \sigma_x,\qquad b = - 4 \sigma_x,\qquad
c = 1 + 6\sigma_x, \qquad
d= - 4 \sigma_x, \qquad e = \sigma_x
\end{equation}%
Similarly,
\begin{equation}
f_i = \alpha_x N_{i - 1}^{n} - 2 \alpha_xN_{i}^{n} + \alpha_x N_{i+1}^{n}
\end{equation}%
\label{eq:matrix_sys}%
\end{subequations}%
where we have taken $\sigma_x = \gamma \Delta t / \Delta x^4$, $\alpha = \Delta t / \Delta x ^2$ and $N^n_i  = (C^3 - C)^n_i$ for brevity. 
This system can then be solved following the methodology presented for cyclic pentadiagonal matrices as found in \cite{navon_pent, gloster2019cupentbatch, gloster2019efficient}, this methodology also influenced our choice of second order accuracy above.
All of the necessary $\vecb$ matrix calculations were carried out using an in-house finite-difference library developed by the authors~\cite{gloster2019custen}.
We use a  time-step size  
\begin{equation}
\Delta t = 0.1 \Delta x,
\label{eq:dt}
\end{equation} 
to ensure stability of the scheme.   This is a heuristic.  Stability of the numerical algorithm~\eqref{eq:1Dnumerical} is not bound by the linear hyperdiffusion term, which is treated implicitly.  Instead, the numerical algorithm may become unstable due to the explicit treatment of the non-linear $C^3$ term.  The instability is therefore controlled by taking the timestep sufficiently small, but still proportional to $\Delta x$, along the lines of Equation~\eqref{eq:dt}.

In the following section we show the convergence features of the scheme which is clearly second order accurate in space due to the choice of the central differences above.

\subsection{1D Convergence Study}
\label{1Dcov}
\begin{figure}[htb]
	\centering
		\includegraphics[width=0.6\textwidth]{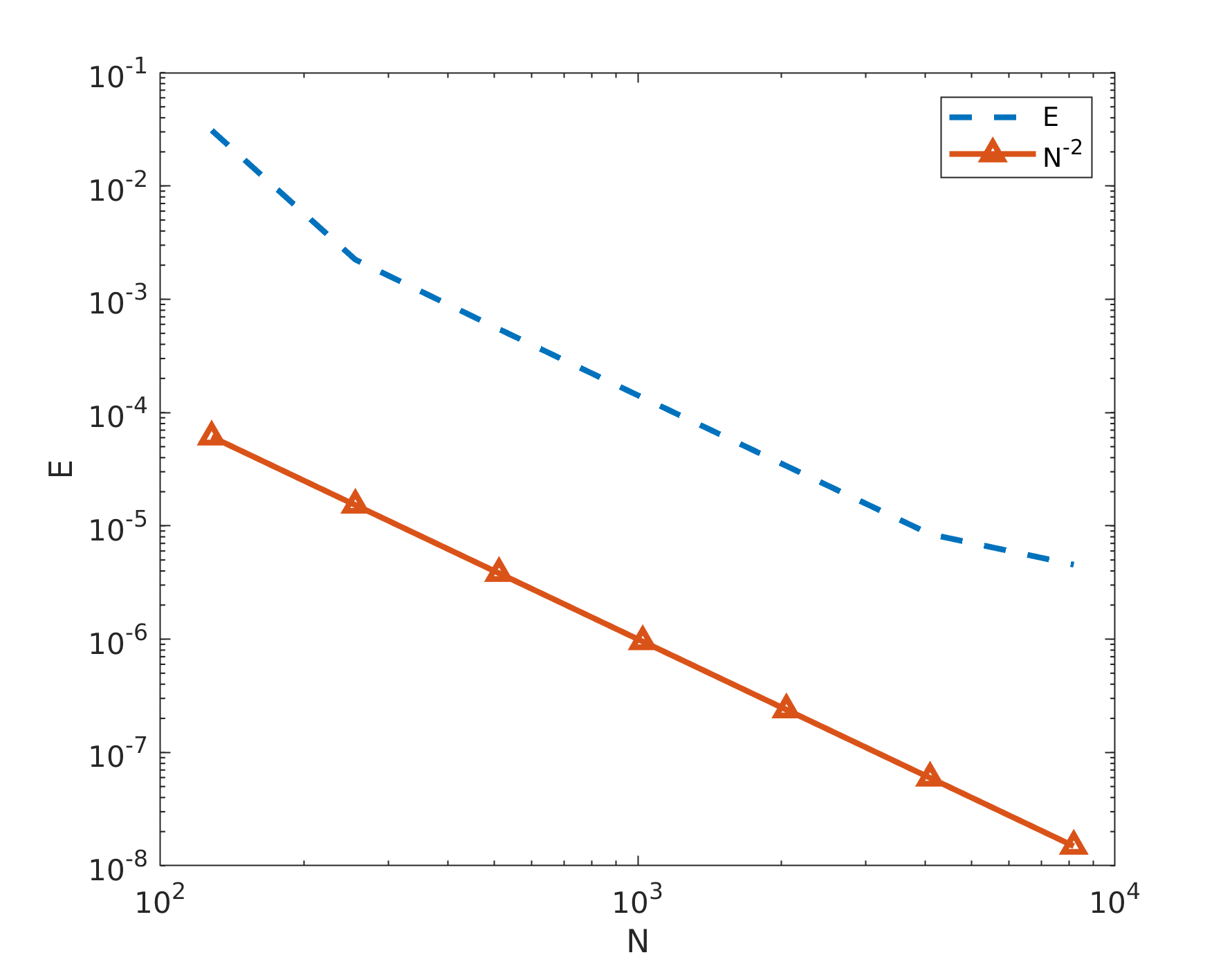}
		\caption{Plot showing the convergence of the scheme in equation~\ref{eq:1Dnumerical}. The slope is $-2$ matching the spatial accuracy of the scheme.}
	\label{convergence1D}
\end{figure}

For the convergence study we set  $\gamma = 0.01$ and $L=2 \pi$.
We set the time-step using Equation~\eqref{eq:dt} and apply a cosine initial condition given by
\begin{equation}
C(x,t=0) = \epsilon \cos(5x),
\end{equation} 
with $\epsilon =  10^{-6}$. 
The convergence results are presented in Figure~\ref{convergence1D} and Table~\ref{table5:converge1D}.
We can see that this scheme for intermediate numbers of grid points is second-order accurate except for very high resolutions where the scheme converges with first-order accuracy.
In all cases, since the solutions of the Cahn--Hilliard equation are smooth, and possess smooth features even on the scale of the transition zones between domains (i.e. no `shocks'), these orders of convergence are acceptable.

\begin{table}
\caption{Details of convergence study of semi-implicit scheme for solving the 1D Cahn--Hilliard equation.}
\centering
\setlength{\tabcolsep}{10pt}
\begin{tabular}{c|c|c} 
\hline
    {$N_x$} & {$E_N$} & {$log_2 (E_N / E_{2N})$} \\
    \hline
    128   & 0.0310      & 3.7932    \\
    256   & 0.0022      & 2.0376    \\
    512   & $5.45 \times 10^{-4}$    & 2.0089    \\
    1024  & $1.35 \times 10^{-4}$    & 2.0022    \\
    2048  & $3.38 \times 10^{-5}$    & 2.0005    \\
    4096  & $8.45 \times 10^{-6}$    & 0.8947	\\
    8192  & $4.54 \times 10^{-6}$    &           \\
\hline
\end{tabular}
\label{table5:converge1D}
\end{table}

\subsection{Batched 1D Scaling}
\label{1Dscale}
We now examine the how the scaling behaviour of the coarsening phenomenon in the 1D Cahn--Hilliard equation. In Reference~\cite{argentina2005coarsening}, theoretical arguments are given for why the size $\typical(t)$ of a typical domain should scale as
\begin{equation}
\typical(t) \propto \log(t).
\label{eq:1DscaleAvg}
\end{equation}
Thus, in order to measure the `typical domain size', we must average across the different domains that appear transiently in a single simulation, and across multiple independent simulations.
To do this, we follow the computational methodology for solving batches of hyperdiffusion equations as presented in~\cite{gloster2019cupentbatch, gloster2019efficient} and combine it with the method for solving individual 1D Cahn--Hilliard equations presented above in Section~\ref{1Dscheme}.
We store each individual $\vecb$ of the system in interleaved format and the cuSten library~\cite{gloster2019custen} is used to compute the required non-linear finite differences.
Each simulation is given its own randomised initial condition $C(x, t=0)$ where values are drawn from a uniform distribution between $-0.1$ and $0.1$.  Thus, the average value of the concentration field is $\langle C\rangle=0$, valid for all times (\textit{cf.} Equation~\eqref{eq:conserved}).  We therefore consider coarsening with respect to the well-mixed state $\langle C\rangle=0$, wherein both components of the binary fluid are present in equal amounts.

The model equation~\eqref{eq:1Dnumerical} is solved in batch mode on a GPU, time-stepping until $T = 100$ and again we set $\gamma = 0.01$.
In order to measure $\typical(t)$ we compute the quantity
\begin{equation}
\ell(t) = \frac{1}{1 - \langle C^2\rangle},\qquad \langle C^2\rangle=\frac{1}{L}\int_0^L [C(x,t)]^2\,\mathd x;
\end{equation}
the connection between the a typical domain size $\ell(t)$ within a given simulation, and the variance $\langle C^2\rangle$ is established in Reference~\cite{LennonAurore}.
The quantity $\ell(t)=(1-\langle C^2\rangle)^{-1}$ is captured for each simulation at every time-step and then averaged across simulations, to produce an average instantaneous value for the typical domain size, which we denote by $\typical(t)$.

\begin{figure}[htb]
    \centering
    \begin{subfigure}[b]{0.49\textwidth}
        \centering
        \includegraphics[width=1.0\textwidth]{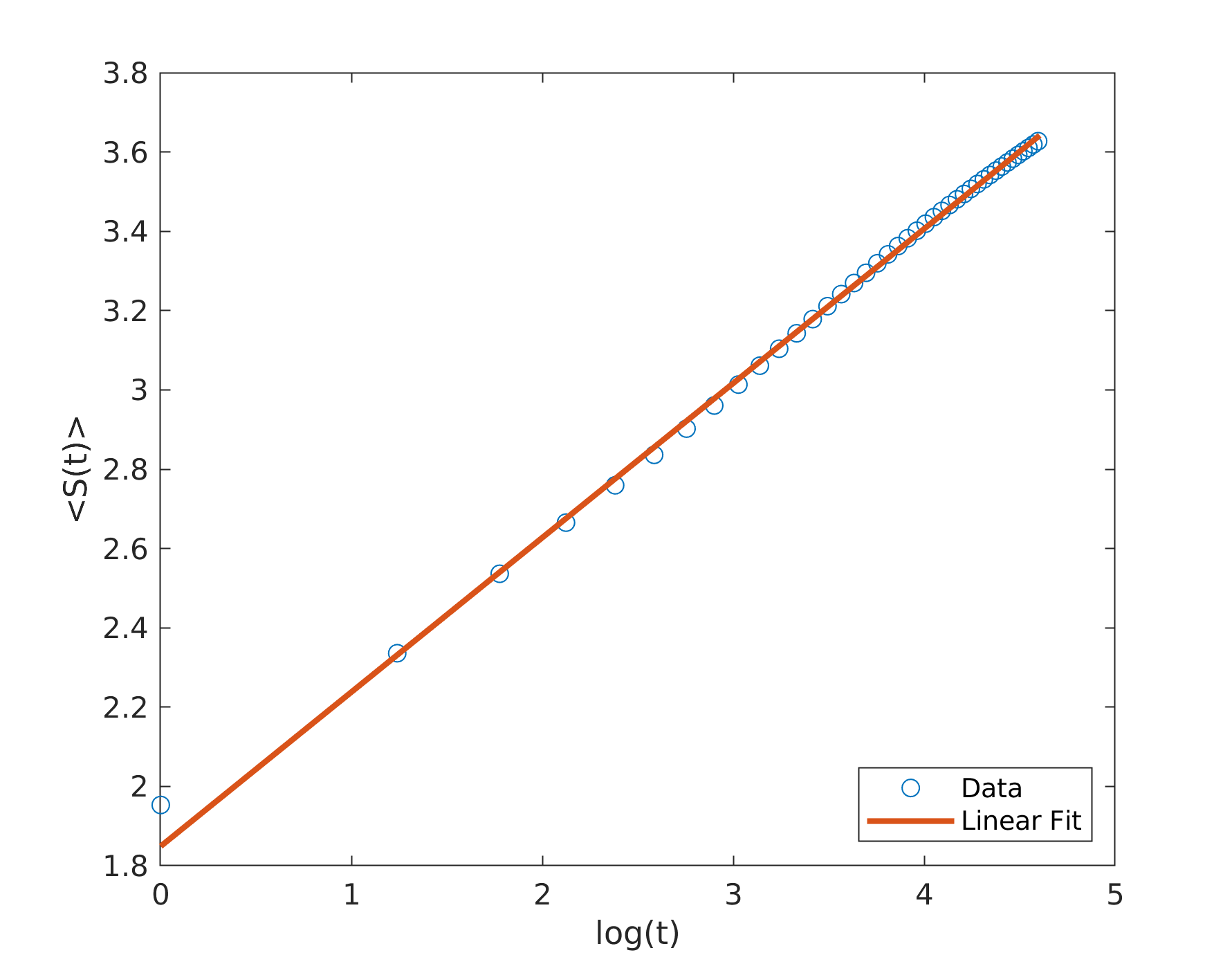}
    \caption{Scaling of domain size $2\pi$.}
    \label{1Dscale2piPlot}
    \end{subfigure}
    \hfill
    \begin{subfigure}[b]{0.49\textwidth}  
        \centering 
        \includegraphics[width=1.0\textwidth]{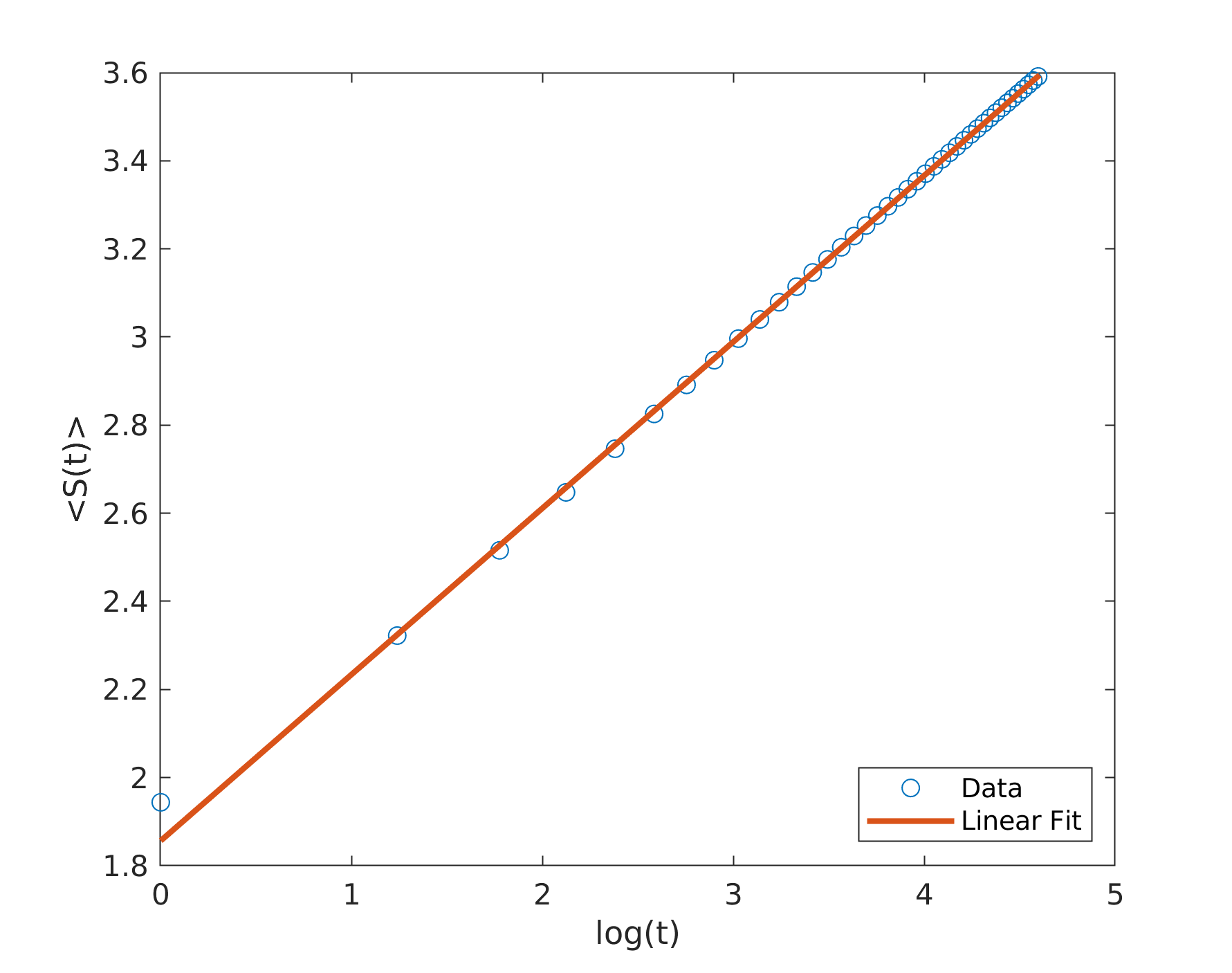}
    \caption{Scaling of domain size $4\pi$.}
    \label{1Dscale4piPlot}
    \end{subfigure}
        \caption{Plots showing the scaling in as a function of $\log{t}$.}
    \label{4pi1024}
\end{figure}

We plot $\typical(t)$ as a function of $\log(t)$ in Figures~\ref{1Dscale2piPlot} and~\ref{1Dscale4piPlot} for domains of size $L=2\pi$ and $L=4\pi$ respectively.
In both cases $\Delta x = 2 \pi / 256$ again to ensure $\Delta x \ll 2\pi \sqrt{\gamma}$ so that the transition region between domains is well resolved. 
In both cases we simulate $2^{20}$ independent systems in parallel on a single GPU in order to ensure the space is well sampled and that we average out the statistic of interest as much as possible. 
In both cases, we fit lines with linear regression as we are testing that the quantities in equation~\eqref{eq:1DscaleAvg} are linearly proportional. 
For the $2\pi$ case in Figure~\ref{1Dscale2piPlot} we report a value of $r = 0.9989$ for the correlation coefficient of the linear regression analysis, while for $4\pi$ in Figure~\ref{1Dscale4piPlot} we report a value of $r = 0.9996$.
Clearly these results confirm the theoretical findings in~\cite{argentina2005coarsening} and provide a good example of how a GPU can be used to efficiently solve a large batch of separate 1D equations.


\begin{figure}[htb]
    \centering
        \includegraphics[width=0.6\textwidth]{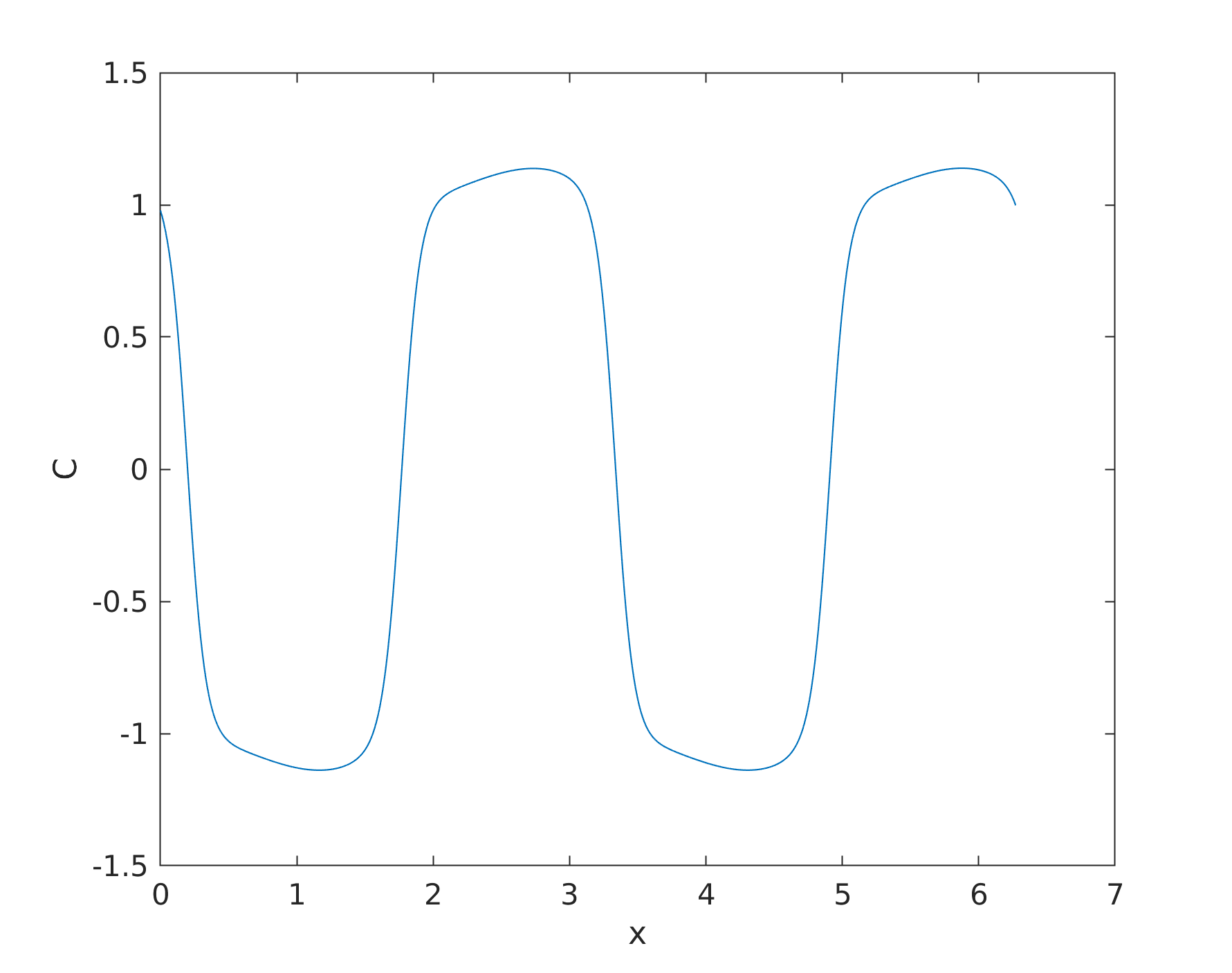}
        \caption{Plot of numerical solution to equation~\eqref{eq:1DcahnForcedTravel} showing regions of $C \approx \pm 1$.}
    \label{fullMinusPlus}
\end{figure}

\section{Forced 1D Cahn--Hilliard Flow Pattern Map}
\label{flowpattern}

In this section we again use the GPU methodology to study the 1D Cahn--Hilliard equation.  The focus of this section is on the forced Cahn--Hilliard equation, which introduces additional parameters into the problem.  The structure of the solution of the Cahn--Hilliard equation then depends on these parameters.  We use a batch of simulations to characterize the parameter space.  The aim here is again twofold: to learn something about the forced Cahn--Hilliard equation, and also to showcase potential uses of the GPU methodology.

To fix ideas, in this section we focus on a travelling-wave forced Cahn--Hilliard equation in one spatial dimension, recalled here as
\begin{equation}
\frac{\partial C}{\partial t} = \frac{\partial^2}{\partial x^2} \left(C^3 - C - \gamma \frac{\partial^2 C}{\partial x^2} \right) + f_0 k \cos(k (x - vt)),\qquad t>0,\qquad x\in (0,L).
\label{eq:1DcahnForced}
\end{equation}
This introduces new parameters into the problem based on the forcing term: the forcing amplitude $f_0$, the wavenumber $k$, and the travelling wave-speed $v$.  In view of the sinusoidal nature of the forcing term, the average concentration $\langle C\rangle$ is conserved and represents a further parameter in the problem.  Assuming a fixed value of $\gamma$, the parameter space to be explored is therefore four-dimensional.

We emphasize that the forced Cahn--Hilliard equation with sinusoidal forcing is a well-established model in the literature, and is used to describe the Ludwig--Soret effect in binary liquids, wherein a temperature gradient induces a concentration gradient~\cite{krekhov2004phase}.  The presence of a travelling-wave forcing has been studied before, e.g. Reference~\cite{weith2009traveling}, and the mathematical properties of the travelling-wave solutions have been characterized in Reference~\cite{lennonWave}.

\begin{figure}[htb]
    \centering
        \includegraphics[width=0.6\textwidth]{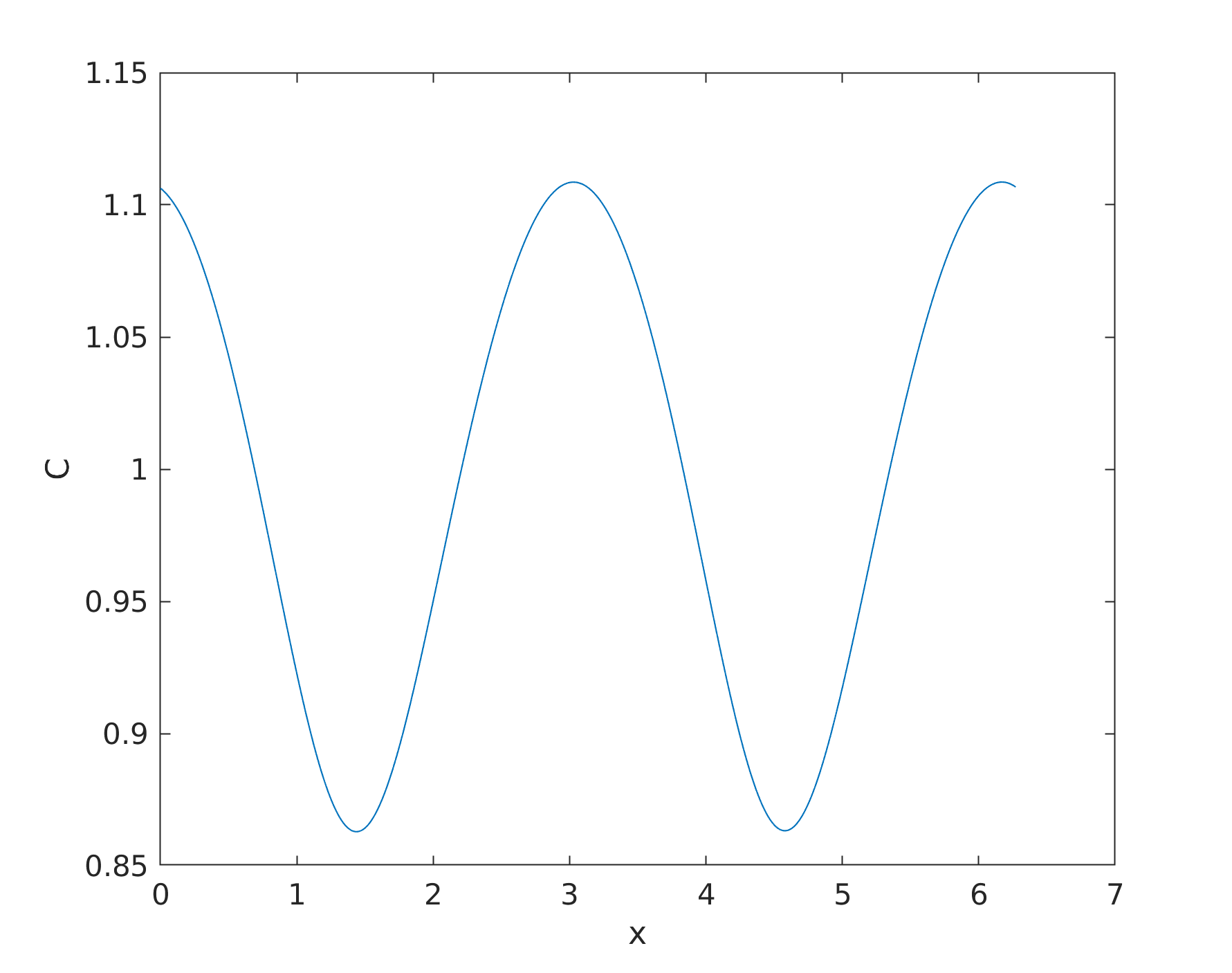}
        \caption{Plot of numerical solution to equation\eqref{eq:1DcahnForcedTravel} showing regions oscillation around a mean value of $C \approx 1$.}
    \label{oscillation}
\end{figure}

Based on Equation~\eqref{eq:1DcahnForced}, three distinct solution types have been isolated previously, by theoretical analysis~\cite{weith2009traveling,lennonWave} (the nomenclature is taken from these references):
\begin{itemize}
\item A2-travelling-wave solutions -- these are travelling-wave solutions of
Equation~\eqref{eq:1DcahnForced}, which consist of a sinusoidal travelling wave superimposed on a mean concentration value $\langle C\rangle$.  Example: Figure~\ref{oscillation}.
\item A1-travelling-wave solutions -- these are travelling-wave solutions of Equation~\eqref{eq:1DcahnForced} which consist of interconnected domains (varying between $C=-1$ and $C=+1$).  The interconnected domain structure moves with the travelling-wave velocity.  
\item A0 solutions -- these occur in regions of the parameter space where the forcing term is not strong enough to impose a travelling-wave structure on the solutions.  Instead, the solution consists of standard domain structures, which move from time-to-time in an irrgular fashion due to the forcing term.  In short, these are intermediate solutions, where there is dynamic competition between phase separation and forcing.
\end{itemize}
Previously, in Reference~\cite{lennonWave}, these different solution types were identified with specific regions of parameter space -- this was done by theoretical analysis in the first instance, combined with a visual inspection of a limited number of numerical simulations.  The result was colloquially called a `flow-pattern map', in analogy with the flow-pattern maps for multiphase flow in the Chemical Engineering literature~\cite{Hewitt1982}.  The aim in this section of the paper is to see if these results can be reproduced using batch solution of Equation~\eqref{eq:1DcahnForced}, together with standard techniques from machine learning.

To produce a good quality flow-pattern map with well-defined regions of parameter space as in Reference~\cite{lennonWave}, a large number of simulations is required followed by some visual inspection.
This approach is slow and inefficient as it leaves the researcher having to study each numerical solution to classify the finding, particularly in cases where an analytic metric cannot be used to help determine the classification of a given solution in the flow-pattern map.
Another key drawback of this method is it can be hard to draw out the precise boundaries between two solution regions, particularly when there is a high number of simulations and a lack of an analytic expression to divide the parameter space.
In this section of the paper, we therefore present a proposed solution which addresses these issues.  The solution has two parts:
\begin{itemize}
\item  First, a method to produce the necessary large batches of simulation in parallel using a GPU, this allows for extremely large and diverse data-sets to be produced for the study, 
\item  Next, an application of k-means clustering to classify the results automatically to produce the desired flow-pattern maps. 
\end{itemize}

Before proceeding with the analysis we first of all rewrite Equation~\eqref{eq:1DcahnForced} in the frame of reference of the travelling wave, this ensures we have a consistent position relative to the wave when feeding it into our classification method.  The desired coordinate transformation is given by
\begin{equation}
\eta = x - vt.
\end{equation}
Under this transformation equation~\eqref{eq:1DcahnForced} becomes 
\begin{equation}
\frac{\partial C}{\partial t} - v \frac{\partial C}{\partial \eta} =  \frac{\partial^2}{\partial \eta^2}\left(C^3 - C - \gamma \frac{\partial^2 C}{\partial \eta^2} \right) + f_0 k \cos(k \eta)
\label{eq:1DcahnForcedTravel}
\end{equation}
%


\begin{figure}[htb]
    \centering
        \includegraphics[width=0.6\textwidth]{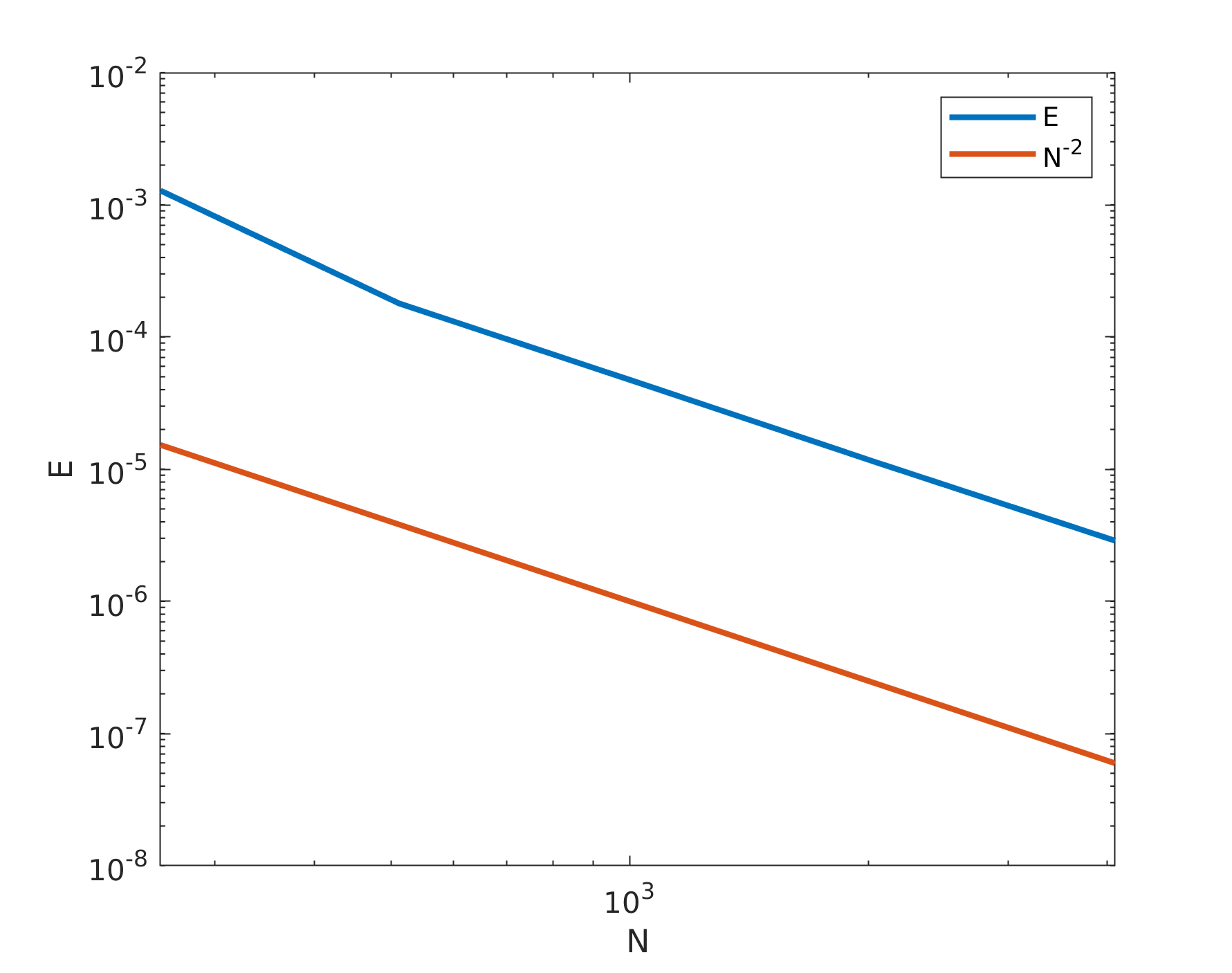}
        \caption{Plot showing the convergence of the forced Cahn--Hilliard scheme in equation~\ref{eq:travellingNumerical}. The slope is $-2$ matching the spatial accuracy of the scheme.}
    \label{convergenceTravel1D}
\end{figure}

\subsection{Numerical Scheme and Convergence Study}
\label{schemeTravel}

Before presenting the analysis of results for equation~\eqref{eq:1DcahnForcedTravel}, we first of all present the numerical scheme and a convergence study.
As in previous sections, we discretise the hyperdiffusion term in Equation~\eqref{eq:1DcahnForcedTravel} implicitly to give a version of equation~\eqref{eq:1Dnumerical} with additional terms on the $\vecb$:
\begin{equation}
C^{n+1} + \Delta t \gamma \frac{\partial^4 C^{n+1}}{\partial \eta^4} =  C^n + \Delta t \left[\frac{\partial^2}{\partial \eta^2} \left(C^3 - C \right)^n + v \frac{\partial C^n}{\partial \eta} + f_0 k \cos(k \eta) \right].
\label{eq:travellingNumerical}
\end{equation}
Thus the solution methodology with pentadiagonal inversions and central differences is the same as Section~\ref{1Dscheme} except we need to discretise the advection term $v (\partial C / \partial \eta)$.
In order to deal with this term we use a standard Hamilton--Jacobi WENO scheme presented in~\cite{fedkiwBook}.
We choose this scheme as it correctly differences the advection term in the direction of travel and is $O(\Delta x^5)$ accurate in smooth regions, this scheme has been implemented using the cuSten library~\cite{gloster2019custen}.


We proceed with an identical convergence study to Section~\ref{1Dcov}, we have set the additional necessary parameters as follows $k = 1$, $v =0.5$ and $f_0 = 1.0$.
The convergence results are presented in Figure~\ref{convergenceTravel1D} and Table~\ref{table5:convergeTravel1D}.
In both cases we can see clear second order convergence as expected as we have been using at least second order accurate stencils in all of our calculations.

\begin{table}
\centering
\caption{Details of convergence study of semi-implicit scheme for solving the 1D forced Cahn--Hilliard equation.}
\setlength{\tabcolsep}{10pt}
\begin{tabular}{c|c|c} 
\hline
    {$N_x$} & {$E_N$} & {$log_2 (E_N / E_{2N})$} \\
\hline
    256   & 0.0013                  & 2.8363    \\
    512   & $1.8 \times 10^{-4}$      & 1.9869    \\
    1024  & $4.54 \times 10^{-5}$  & 2.0086    \\
    2048  & $1.13 \times 10^{-5}$  & 1.9673    \\
    4096  & $2.88 \times 10^{-5}$  &           \\
\hline
\end{tabular}
\label{table5:convergeTravel1D}
\end{table}

\subsection{k-Means Clustering}
\label{kmeans}
k-means clustering is a standard method for partitioning observations into clusters based on distance to the nearest centroid.
Distance in this sense is calculated using Euclidean distance in a space of suitable dimension for the dataset. 
In this work we use MATLAB's implementation of the k-means clustering algorithm which is itself based on Lloyd's algorithm~\cite{kmeans}.
This algorithm is built on two steps which alternate, the first step is to assign each observation to a cluster based on its Euclidean distance to each of the centroids, the assigned cluster is the closest centroid.
Then the second step is to calculate a new set of centroids by calculating the mean position of all the observations within each assigned cluster. 
These steps alternate until convergence has been reached.
The algorithm is initialised by randomly creating the desired number of centroid (3 in our case).

\subsection{Clustering Results}
\label{kmeanResults}
In this section we present the results of running batches of the Cahn--Hilliard equation with a travelling wave and classifying the solutions using the k-means algorithm.
The methodology will be to solve batches of equations where we fix the wave-speed $v$ and wave-number $k$ for all systems within a given batch and then vary $\langle C \rangle$ and $f_0$ within the batch.
$\langle C \rangle$ will be varied in the range $[0,1.5]$ and $f_0$ in the range $[0,2]$.
In order to accurately capture the travelling wave we set the domain size to $2\pi$ and set the number of points in the domain to $N = 512$, it was found that this high number of $N$ was required in order to accurately resolve the travelling wave at high values of $\langle C \rangle$ and $f_0$.
This choice of $N$ also gives us the maximum batch size of the simulations that can be run on a single GPU, in this study our GPU has 12GB of space, thus we determine that we can have a total number of 589824 simulations in each batch.
Thus the resolution of our flow pattern maps will be $768 \times 768$ dividing the ranges of $f_0$ and $\langle C \rangle$ appropriately.
The initial conditions in this section are drawn from a random uniform distribution of values between $-0.1$ and $0.1$ with $\gamma$ set to a value of $0.01$.
We simulate all of the systems up to a final time of $T = 50$.

\begin{figure}[htb]
    \centering
        \includegraphics[width=0.6\textwidth]{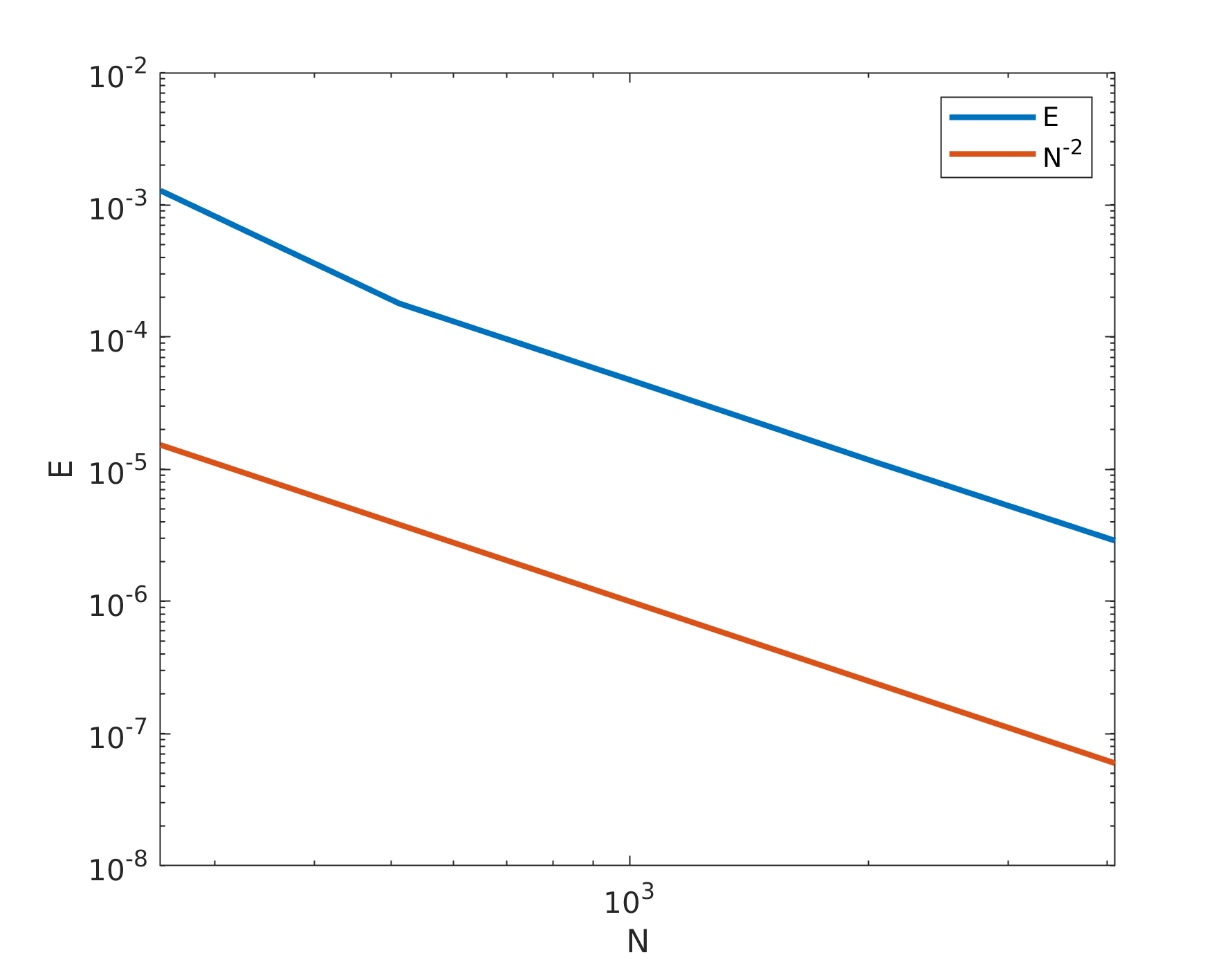}
        \caption{Flow-pattern map of equation~\eqref{eq:travellingNumerical} with  $k = 1.0$ and $v = 0.5$, the white area corresponds to A0 solutions, grey to A1 solutions and black to A2 solutions.}
    \label{plot1}
\end{figure}

We begin by examining the case $k = 1.0$ and $v = 0.5$, the results of which are presented in Figure~\ref{plot1}.
In this plot we can see the clear extraction of regions of different solutions, the top left grey region corresponds to solution type A1 and the right side black region corresponds to solution type A2.
A mix of solutions can be seen in the bottom left white region of the domain corresponding to region A0 along with some areas of A1 and A2.
Between these two regions k-means has been able to pick out an exact boundary extending from $f_0 \approx 0.9$ to the top of the domain.
Some noise can be seen along this boundary where the k-means algorithm has struggled to differentiate between very similar solution types as the behaviour transitions,
This also explains the less well behaved bottom left region where the algorithm has been unable to deal with areas near the boundaries.

The results presented in Figure~\ref{plot1} are in keeping with those presented in~\cite{lennonWave}; where the classification was performed by visual inspection.
Three regions in approximately the same locations were found, solution type A0 in the bottom left, A1 on the top left and then A2 on the right side of the parameter space. 
We now increase the value of $v$ to see if the boundary between A0 and A1 moves higher up the $f_0$ axis.

\begin{figure}[htb]
    \centering
        \includegraphics[width=0.5\textwidth]{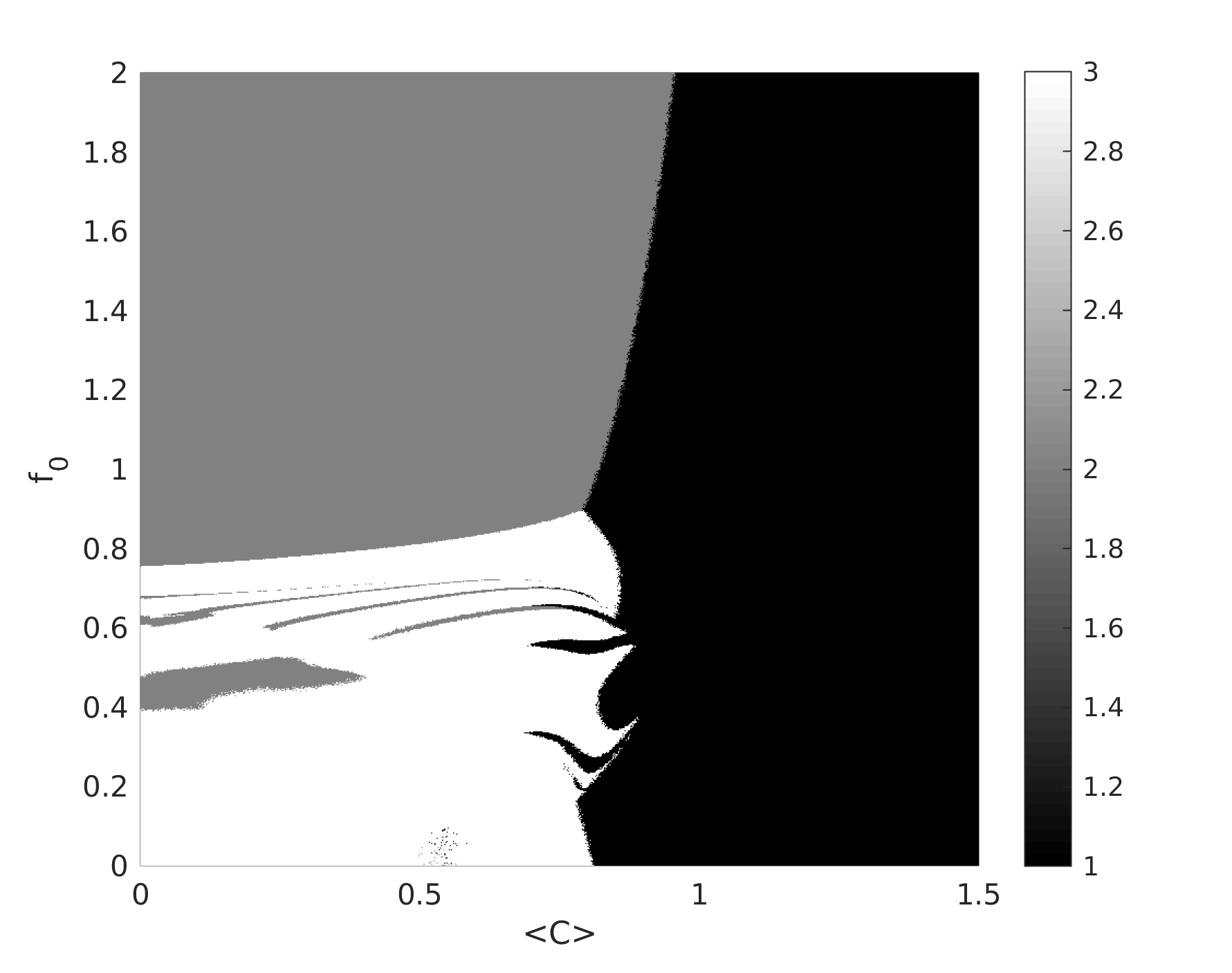}
        \caption{Flow-pattern map of equation~\eqref{eq:travellingNumerical} with  $k = 1.0$ and $v = 1.0$, the white area corresponds to A0 solutions, grey to A1 solutions and black to A2 solutions.}
    \label{plot2}
\end{figure}

We now examine the case $k = 1.0$ and $v = 1.0$, the results of which are presented in Figure~\ref{plot2}.
The k-means algorithm has struggled here to extract clean boundaries between the A0 region and the other two. 
This can be attributed to the fact that these are hard to distinguish, even by eye.
We can see that the algorithm has successfully classified the regions A1 and A2 broadly quite well, with a less well defined boundary than previously.
Thus we can conclude that k-means is a useful approach for broadly classifying large parameter spaces in terms of flow pattern maps yet will struggle with more subtle transitions between regions.

Instead k-means could be used as a first step in building a model to classify simulation data for a flow pattern map.
Indeed one could take the well defined regions from a large number of simulations to then train a logistic regression classifier or neural network, these may produce better results as they may detect features that are not as obvious to the eye or representable using a Euclidean distance as in the k-means algorithm.
Also additional features beyond the raw solution could be included in the dataset such as a time series of the free energy or an increased number of time-steps of given solutions.


\section{Conclusions}
\label{con}

Summarizing, we have presented new algorithms that utilise improved data access when solving batches of Triadiagonal and Pentadiagonal matrices that all share the same LHS matrix. This new methodology of an interleaved batch of RHS matrices along with a single LHS achieves large reductions in data usage along with a substantial speed-up when compared with the existing state of the art. The reduction in data usage allows for greater use of hardware resources to be made with significant extra space that can now be devoted to solving
more systems of equations simultaneously rather than unnecessarily storing data. The extra
equations now being solved by one GPU along with the speed-ups provided by the new
implementation are a significant improvement over the state of the art in applications where
only one LHS matrix is required for the solution of all the systems in the batch. In addition the reduction in the number of GPUs one would need when solving very large batches,
coupled with the speed-ups achieved, leads to a reduction in electricity usage which has both economical and environmental implications for HPC.

In this article, we have used a non-trivial physical problem to motivate the development of the GPU methodology -- notably, the Cahn--Hilliard equation in one spatial dimension.  We have demonstrated how the GPU methodology can be used to perform up to $2^{20}$ simulations of the Cahn--Hilliard equation -- the resulting data can be used to build a statistically robust description of the coarsening phenomenon in one-dimensional phase separation.  The sheer number of simulations used here to reveal the $\overline{\ell}(t)\sim \log (t)$ scaling behaviour in one sptial dimension is without precedent in the literature.

We have also used the physical test-bed of the Cahn--Hilliard equation with a forcing term to show how the different solution-types of equation can be classified as a function of the forcing-term parameters.  Again, we use a very large number of simulations to generate simulation data which in this instance can be fed into standard machine-learning algorithms.    This approach automates the construction of parameter spaces (`flow-pattern maps') which were previously produced by hand -- although visual inspection is still required on occasion for ambiguous edge cases.

We emphasize finally that the numerical algorithms at the heart of the Cahn--Hilliard test-bed are rather generic (finite-differences, batch solution of partial differential equations, etc.).  Consequently, the approach outlined in this article may find broad use in Computational Fluid Mechanics, and in Computational Science and Engineering more broadly.


\section*{Acknowledgements}
Andrew Gloster acknowledges funding received from the UCD Research Demonstratorship.   
Enda Caroll acknowledges funding recieved under the Government of Ireland Postgraduate Scholarship Programme funded by Irish Research Council, grant number GOIPG/2018/2653.
All authors gratefully acknowledge the support of NVIDIA Corporation with the donation of the Titan X Pascal GPUs used for this research.  
The authors also thank Lung Sheng Chien for helpful discussions throughout this project.
The authors wish to acknowledge the DJEI/DES/SFI/HEA Irish Centre for High-End Computing (ICHEC) for the provision of computational facilities and support.

\newpage
\appendix

\section{Tridiagonal Algorithm}

\begin{algorithm}[H]
\caption{Pre-Factorization Step of the Thomas Algorithm (performed on CPU)}
\begin{algorithmic}[1]
	\Function {preFactConstantTri}{$\mathbf{a}$, $\mathbf{b}$, $\mathbf{c}$, $N$}
	\State $c_1 \leftarrow c_1 / b_1$
	\For{$i \leftarrow 2, n - 1$}
    	\State $b_i \leftarrow b_i - a_i c_{i - 1}$
    	\State $c_i \leftarrow c_i / b_i$ 
    \EndFor
    \State $b_N \leftarrow b_N - a_N c_{N - 1}$
    \EndFunction
    \label{algo:preFactTri}
\end{algorithmic}
\end{algorithm}

\begin{algorithm}[H]
\caption{Batch Thomas Algorithm (performed on GPU)}
\begin{algorithmic}[1]
	\Function {cuThomasConstantBatch}{$\mathbf{a}$, $\mathbf{b}$, $\mathbf{c}$, $\mathbf{d}$, $N$, $M$}
	\State rowID $\leftarrow$ threadId.x + blockDim.x * blockIdx.x
	\If {rowID $< M$}
		\State $d_{rowID} \leftarrow d_{rowID} / b_1$
		\For{$i \leftarrow 2, n$}
			\State rowID $\leftarrow$ rowID$ + M$
	    	\State $d_{rowID} \leftarrow (d_{rowID} - a_i d_{rowID - M}) / b_i$
	    \EndFor
		\For{$i \leftarrow n - 1, 1$}
			\State rowID $\leftarrow$ rowID$ - M$
	    	\State $d_{rowID} \leftarrow d_{rowID} - c_i d_{rowID + M}$
	    \EndFor
	\EndIf
    \EndFunction
    \label{algo:batchThomas}
\end{algorithmic}
\end{algorithm}

\newpage
\section{Pentadiagonal Algorithm}

\begin{algorithm}[H]
\caption{Pre-Factorization Step of the Pentadiagonal Inversion Algorithm (performed on CPU)}
\begin{algorithmic}[1]
	\Function {preFactConstantPent}{$\mathbf{a}$, $\mathbf{b}$, $\mathbf{c}$, $\mathbf{d}$, $\mathbf{e}$, $N$}
	\State $d_1 \leftarrow d_1 / c_1$
	\State $e_1 \leftarrow e_1 / c_1$

	\State $c_2 \leftarrow c_2  - b_2 d_1$
	\State $d_2 \leftarrow (d_2 - b_2 e_1) / c_2$
	\State $e_2 \leftarrow e_2 / c_2$

	\For{$i \leftarrow 3, n - 2$}
    	\State $b_i \leftarrow b_i - a_i d_{i - 2}$
    	\State $c_i \leftarrow c_i - a_i c_{i - 2} - b_i d_{i - 1}$ 
    	\State $e_i \leftarrow e_i / c_{i}$
    	\State $d_i \leftarrow (d_i - b_i e_{i - 1}/ c_i$ 
    \EndFor
    \State $b_{N - 1} \leftarrow b_{N - 1} - a_{N - 1}d_{N - 3}$
    \State $c_{N - 1} \leftarrow c_{N - 1}  - a_{N - 1}e_{N-3} - b_{N - 1} d_{N - 2}$
	\State $d_{N - 1} \leftarrow (d_{N - 1} - b_{N - 1} e_{N - 2}) / c_{N - 1}$

	\State $b_N \leftarrow b_N - a_N d_N$
	\State $c_N \leftarrow c_N - a_N e_{N - 2} - b_{N} d_{N -1}$
    \EndFunction
    \label{algo:preFactPent}
\end{algorithmic}
\end{algorithm}

\begin{algorithm}[H]
\caption{Batch Pentadiagonal Inversion Algorithm (performed on GPU)}
\begin{algorithmic}[1]
	\Function {cuPentConstantBatch}{$\mathbf{a}$, $\mathbf{b}$, $\mathbf{c}$, $\mathbf{d}$, $\mathbf{e}$, $\mathbf{f}$, $N$, $M$}
	\State rowID $\leftarrow$ threadId.x + blockDim.x * blockIdx.x
	\If {rowID $< M$}
		\State $f_{rowID} \leftarrow f_{rowID} / c_1$
		\State rowID $\leftarrow$ rowID$ + M$
		\State $f_{rowID} \leftarrow (f_{rowID} - b_2 f_{rowID - M}) / c_2$
		\For{$i \leftarrow 3, n$}
			\State rowID $\leftarrow$ rowID$ + M$
	    	\State $f_{rowID} \leftarrow (f_{rowID} - a_i f_{rowID - 2 M} - b_i f_{rowID - M} ) / c_i$
	    \EndFor
	    \State rowID $\leftarrow$ rowID$ - M$
	    \State $f_{rowID} \leftarrow f_{rowID} - d_{N - 1} f_{rowID + M}$
		\For{$i \leftarrow n - 2, 1$}
			\State rowID $\leftarrow$ rowID$ - M$
	    	\State $f_{rowID} \leftarrow f_{rowID} - d_i d_{rowID + M} - e_i f_{rowID + 2 M}$
	    \EndFor
	\EndIf
    \EndFunction
    \label{algo:batchPent}
\end{algorithmic}
\end{algorithm}

\section*{References}


\end{document}